\documentclass[11pt]{article}
\usepackage[margin=1in]{geometry}

\usepackage{bbm}	
\usepackage{amsthm,amssymb,amsmath}
\usepackage{thmtools,thm-restate}
\usepackage{tikz}
\usetikzlibrary{arrows,automata}
\usepackage{multirow}
\usepackage{tabularx}
\usepackage[ruled,lined,boxed]{algorithm2e}
\usepackage{algorithmic}
\usepackage{comment}
\usepackage[caption=false]{subfig}

\usepackage[mathic]{mathtools}
\usepackage{algorithmic}
\usepackage[T1]{fontenc}
\usepackage{bbm}
\usepackage{float}
\usepackage{booktabs}
\usepackage{libertine}
\usepackage{authblk}
\newtheorem{proposition}{Proposition}
\newtheorem{exmp}{Example}

\DeclarePairedDelimiter\ceil{\lceil}{\rceil}

\usepackage{thm-restate}
\usepackage[libertine]{newtxmath} % must load after ams packages
\usepackage[scaled=0.96]{zi4} % use Inconsolata as the typewriter font

\usepackage{a4wide, dsfont, microtype, xcolor, paralist}

\usepackage[ocgcolorlinks]{hyperref} % Option ocgcolorlinks makes links only colored on screen and not on printouts
\colorlet{DarkRed}{red!50!black}
\colorlet{DarkGreen}{green!50!black}
\colorlet{DarkBlue}{blue!50!black}
\hypersetup{
	linkcolor = DarkRed,
	citecolor = DarkGreen,
	urlcolor = DarkBlue,
	bookmarksnumbered = true,
	linktocpage = true
}

\let\epsilon\varepsilon

\usepackage{xspace}

\usepackage{tikz}
\usetikzlibrary{decorations.pathreplacing}
\usetikzlibrary{plotmarks}
\usetikzlibrary{positioning,automata,arrows}
\usetikzlibrary{shapes.geometric}
\usetikzlibrary{decorations.markings}
\usetikzlibrary{positioning, shapes, arrows}

\PassOptionsToPackage{usenames,dvipsnames,svgnames}{xcolor}
\definecolor{orange}{RGB}{235,90,0}
\definecolor{darkorange}{RGB}{175,30,0}
\definecolor{turkis}{RGB}{131,182,182}
\definecolor{darkturkis}{RGB}{31,82,82}
\definecolor{green}{RGB}{102,180,0}
\definecolor{darkgreen}{RGB}{51,90,0}
\definecolor{myblue}{RGB}{0,0,213}
\definecolor{mydarkblue}{RGB}{0,0,100}
\definecolor{mybrightblue}{HTML}{74B0E4}
\definecolor{mybrighterblue}{HTML}{B3EAFA}
\definecolor{lila}{RGB}{102,0,102}
\definecolor{darkred}{RGB}{139,0,0}
\definecolor{darkyellow}{RGB}{188,135,2}
\definecolor{brightgray}{RGB}{200,200,200}
\definecolor{darkgray}{RGB}{50,50,50}
\definecolor{amaranth}{rgb}{0.9, 0.17, 0.31}
\definecolor{alizarin}{rgb}{0.82, 0.1, 0.26}
\definecolor{amber}{rgb}{1.0, 0.75, 0.0}
\definecolor{green(ryb)}{rgb}{0.4, 0.69, 0.2}
\definecolor{hanblue}{rgb}{0.27, 0.42, 0.81}
\definecolor{grannysmithapple}{rgb}{0.66, 0.89, 0.63}

\newtheorem{theorem}{Theorem}[section]
\newtheorem{lemma}[theorem]{Lemma}

\newtheorem{lemma-rstbl}[theorem]{Lemma}
\newtheorem{obs-rstbl}[theorem]{Observation}
\newtheorem{theorem-rstbl}[theorem]{Theorem}

\usepackage{environ}
\NewEnviron{cproblem}[1]{%
\begin{center}\fbox{\parbox{5.5in}{%
    {\centering\scshape #1\par}%
    \parskip=1ex
    \everypar{\hangindent=1em}%
    \BODY
}}\end{center}}

\title{When Can Liquid Democracy Unveil the Truth ?}

\author[1]{Ruben Becker}
\author[1]{Gianlorenzo D'Angelo}
\author[1]{Esmaeil Delfaraz}
\author[2]{Hugo Gilbert}
\affil[1]{\normalsize Gran Sasso Science Institute, L'Aquila, Italy}
\affil[2]{Université Paris-Dauphine, Université PSL, CNRS, LAMSADE, 75016 Paris, France}

\date{}
\begin{document}
\maketitle
\begin{abstract}
    Liquid democracy is a voting paradigm that allows voters that are part of a social network to either vote directly or delegate their voting rights to one of their neighbors. The delegations are transitive in the sense, that a voter who decides to delegate, delegates both her own vote and the ones she has received through delegations.
    The additional flexibility of the paradigm allows to transfer voting power towards a subset of voters ideally containing the most expert voters on the question at hand. 
    It is thus tempting to assume that liquid democracy can lead to more accurate decisions. This claim has been investigated recently, maybe most importantly, by Kahng, Mackenzie, and Procaccia~\cite{kahng2018liquid} and 
    Caragiannis and Micha~\cite{caragianniscontribution}, who provide however mostly negative results using a model similar to the uncertain dichotomous choice model.
    
    In this paper, we provide new insights on this question. 
    In particular, we investigate the so-called \textit{ODP}-problem that has been formulated by Caragiannis and Micha~\cite{caragianniscontribution}. Here, we are in a setting with two election alternatives out of which one is assumed to be correct. In \textit{ODP}, the goal is to organise the delegations in the social network in order to maximize the probability that the correct alternative, referred to as ground truth, is elected. While the problem is known to be computationally hard, we strengthen existing hardness results by providing a novel strong approximation hardness result:
    For any positive constant $C$, we prove that, unless $P=NP$, there is no polynomial-time algorithm for \textit{ODP} that achieves an approximation guarantee of $\alpha \ge (\ln n)^{-C}$, where $n$ is the number of voters. 
    The reduction designed for this result uses poorly connected social networks in which some voters suffer from misinformation. Interestingly, under some hypothesis on either the accuracies of voters or the connectivity of the network, we obtain a polynomial-time $1/2$-approximation algorithm. 
    This observation proves formally that the connectivity of the social network is a key feature for the efficiency of the liquid democracy paradigm.
    Lastly, we run extensive simulations and observe that simple algorithms (working either in a centralized or decentralized way) 
    outperform direct democracy on a large class of instances. 
    Overall, our contributions yield new insights on the question in which situations liquid democracy can be beneficial.
\end{abstract}
% \thispagestyle{empty}
% \pagebreak

\section{Introduction}\label{secIntro}

\emph{Liquid Democracy} (LD) is a recent voting paradigm which aims to modernize the way we make collective decisions~\cite{blum2016liquid, green2005direct}. 
It relies on modern tools from the Internet, as social networks, to make democracy more flexible, interactive and accurate~\cite{brill2018interactive}. 
In a nutshell, LD allows voters to delegate transitively along a social network. 
Indeed, each voter may decide to vote directly or to delegate her vote to one of her neighbors. 
This neighbor can in turn delegate her vote and the votes that have been delegated to her to someone else. 
As a result, these delegations will flow until they reach a voter who decides to vote. 
This voter is called the \emph{guru} of the people she represents and has a voting weight equal to the number of people who directly or indirectly delegated to her, including her own.  
LD is implemented in several online tools~\cite{behrens2014principles, hainisch2016civicracy,hardt2015google, paulin2010vzupa} and has been used by several political parties (e.g., the German Pirate party) for inner-decision making.\footnote{A recent survey~\cite{paulin2020overview} reviews the different tools that have implemented liquid democracy features as well as the organizations that have supported its use.} 
The framework is praised for its flexibility, as it enables voters to vote directly for issues on which they feel both concerned and expert and to delegate for others. 
As a result, LD offers a middle-ground between direct and representative democracy and hopefully provides the best of both worlds.

Notably, as delegations can be motivated by the will to find a more expert representative than oneself, LD should concentrate the voting power in the hands of the most expert voters. 
At first glance, this seems as a desirable feature in particular when the election aims to discover a \emph{ground truth} (i.e., one of the alternative is the correct one to elect). 
Indeed, gurus should be on average more informed and hence more likely to vote for the ground truth. 
This claim was previously investigated by Kahng, Mackenzie and Procaccia~\cite{kahng2018liquid} and Caragiannis and Micha~\cite{caragianniscontribution} who mostly provided negative results. 
We proceed by reporting on their results in more detail.

Khang, Mackenzie and Procaccia~\cite{kahng2018liquid} study the following simple model. 
The election has only two alternatives: a correct alternative $T$ and an incorrect one $F$. 
Voters are nodes in a social network, modeled by a directed graph (i.e., there is an arc from voter $i$ to voter $j$ if voter $i$ knows $j$). 
Each voter $i$ has an accuracy parameter $p_i$ associated to her that indicates how well-informed she is: 
if voter $i$ chooses to vote, she will cast a vote for $T$ with probability $p_i$ and a vote for $F$ with probability $1-p_i$.\footnote{This model is often referred to as the uncertain dichotomous choice model.} 
Voters can delegate their vote to any neighbor that they approve and this delegation can be redelegated transitively. 
Each voter who decides to vote is hence weighted by the number of people she represents.  
Lastly, the outcome of the election is decided using weighted majority. 

The authors study local delegation mechanisms, which prescribe a behavior (i.e., delegating and to whom or voting) to each voter using as information only the voter's neighborhood. To evaluate them, they define the gain of a mechanism w.r.t.\ a given instance as the increase (or decrease if this quantity is negative) in the probability that it makes a correct decision w.r.t.\ direct voting on this instance. The authors showed that, under their model, there is no local mechanism satisfying both \emph{positive gain} (there are some large instances in which the mechanism has positive gain bounded away from 0) and \emph{do no harm} (the possible loss of the mechanism goes to 0 as the number of voters grows). However, the authors proposed a slightly more centralized mechanism, called {\tt greedyCap}, which achieves both of these two objectives.  

Under a slightly different model, Caragiannis and Micha~\cite{caragianniscontribution}, confirmed that local mechanisms could be very inefficient (their observations use instances where any local mechanism can be outperformed either by direct democracy or a dictatorial rule). The authors also investigated a centralized problem, the \emph{Optimal Delegation Problem} (ODP for short), in which one aims to set the delegations in an optimal way and proved that it was  computationally hard to approximate within an additive constant. The authors find both of these results to be surprising and a strong criticism to LD.

\paragraph{Our Contribution.}
In this work we continue to investigate the accuracy of the LD paradigm. 
While some of our results are also negative, our gaze is not particularly severe on liquid democracy. 
Indeed, we believe that a loss of accuracy resulting from concentrating the voting power in too few hands is a  pitfall which has been well understood since the Condorcet majority theorem~\cite{de1785essai}. 
Moreover, we believe that the hardness of the \emph{optimal delegation problem} is the rule more than the exception for non-trivial graph problems involving probabilities. Our aim here is to provide indications through approximation results and simulations on the type of elections on which LD can be beneficial or problematic. 

On the theoretical side, we prove that, for any positive constant $C$,  unless $P=NP$, there is no polynomial-time algorithm for \textit{ODP} that achieves an approximation guarantee of $\alpha \ge (\ln n)^{-C}$, where $n$ is the number of voters. 
The reduction designed for this result uses poorly connected social networks in which some voters suffer from misinformation. Interestingly, under some hypothesis on either the accuracies of voters or the connectivity of the network, we instead obtain a polynomial-time $1/2$-approximation algorithm. 
This observation proves formally that the connectivity of the social network is a key feature for the efficiency of LD. Lastly, we run extensive simulations and observe that simple algorithms (working either in a centralized or decentralized way) outperform direct democracy on a large class of instances. 

\section{Related Work}\label{secrw}

Studying the accuracy of group-judgmental processes using the uncertain dichotomous choice model is a long-standing line of research in computational social choice~\cite{berend2008effectiveness, de1785essai,grofman1983thirteen, nitzan1982optimal,shapley1984optimizing,terzopoulou2019optimal}. 
One of the most well-known results within this direction is the Condorcet majority theorem~\cite{de1785essai} which asserts that if voters have an accuracy greater than 0.5, then the probability of electing the ground truth using the majority rule will tend towards one as the number of voters increases. 
Numerous other results have followed this pioneering theorem by researchers either investigating variants of this theorem \cite{conitzer2013maximum,estlund1994opinion, grofman1983thirteen}, comparing the accuracy of different rules~\cite{berend2008effectiveness,gradstein1986performance}, or investigating the optimal voting rule within some class (e.g., weighted majority rules)~\cite{nitzan1982optimal,shapley1984optimizing}.

Kahng, Mackenzie and Procaccia~\cite{kahng2018liquid} as well as Caragiannis and Micha~\cite{caragianniscontribution} have investigated the accuracy of the LD framework following this line of research.
These two works that have been detailed in the introduction, are part of a recent trend in the AI literature trying to explore the algorithmic properties of new voting paradigms similar to LD~\cite{abramowitz2018flexible, bloembergen2019rational, boldi2009voting, brill2018interactive, brill2018pairwise, caragianniscontribution, colleysmart, dey2021parameterized, escoffier2018convergence, golz2018fluid, green2005direct, kahng2018liquid, kavitha2020popular, kotsialou2018incentivising, zhang2020power, zhang2021tracking}. 
We now shortly detail different questions addressed by these works.

The question of the expected accuracy of voters (not of the process in itself) has been considered by Green-Armytage~\cite{green2005direct} in a setting where for every voter and issue, a point on the real line represents the opinion of the voter about the issue. The voter's best guess on each issue is the sum of the `true' position of the voter on the issue plus an error term which is modeled as a random variable. Green-Armytage introduces the notion of expressive loss of a voter as the squared distance between her actual vote (i.e., the vote determined by herself directly or her representative indirectly) and her position. He shows that LD outperforms direct democracy in the sense that, in expectation, LD reduces voters' expressive loss. Very recently, Zhang and Grossi~\cite{zhang2021tracking} considered a model very similar to the one studied by Kahng, Mackenzie, and Procaccia~\cite{kahng2018liquid} and Caragiannis and Micha~\cite{caragianniscontribution}. They assumed that voters could delegate a fraction of their weight to their neighbors and proposed a polynomial time algorithm to find a delegation graph that discovers the ground truth with the highest probability on a complete graph. Also, Zhang and Grossi~\cite{zhang2021tracking} proposed new delegation games (a game theoretic model of LD's delegation process), called \textit{greedy delegation games}, where each voter's utility corresponds the expected accuracy of her gurus. They showed that, in such games, Nash Equilibria (NE) correspond to acyclic delegation graphs and and that NE with weighted profiles (i.e., where voters can delegate fractions of their weights) are never worse than NE with pure delegations (where a voter either votes or delegates all of her weight). %Also, they defined the price of anarchy for such delegation games as the worst case ratio between the minimum accuracy under a NE that is obtained in a delegation game, and the accuracy of voters in optimal solution. 

Several articles have warned against a potential pitfall of LD being that some gurus could accumulate an undesirably high voting power~\cite{kahng2018liquid,kling2015voting}. This issue was addressed by Gölz et al.~\cite{golz2018fluid} and Dey, Maiti and Sharma~\cite{dey2021parameterized}. Gölz et al.~\cite{golz2018fluid} adopted a centralized optimization viewpoint to minimize this maximum voting power. The authors provided approximation hardness results as well as approximation algorithms and gave empirical evidence that allowing voters to specify multiple possible delegations could largely decrease the maximum voting power of a guru. Dey, Maiti and Sharma~\cite{dey2021parameterized} considered a problem similar to the one of Gölz et al.~\cite{golz2018fluid}, where one seeks a delegation graph such that each guru receives less delegations than a given parameter $\lambda$. Dey, Maiti and Sharma~\cite{dey2021parameterized} provided some parameterized complexity results for this problem  with respect to parameters $\lambda$, number of sink nodes and the maximum degree of the initial social graph.

Christoff and Grossi~\cite{christoffbinary} studied other possible pitfalls of LD as delegation cycles or possible inconsistencies that can occur if voters should vote on different but connected issues. Indeed, if these issues are connected, then voters may violate some rationality constraint by delegating their votes to different (disagreeing) gurus. 
For instance, Brill and Talmon~\cite{brill2018pairwise} have investigated a setting called \textit{pairwise liquid democracy}, in which each voter should provide a complete ranking over candidates. To do so, each voter may delegate different binary preference queries to different representatives. In this case, the delegation process may yield incomplete or intransitive preferences. The authors discuss several methods to complete such ballots or to proceed with the vote.

This illustrates the difficulty of unraveling ballots arising from complex delegation processes, a problem that was tackled by Colley, Grandi and Novaro~\cite{colleysmart}, who designed several unravelling procedures that may be used with delegation processes even more general than the one of LD. Similarly to Kotsialou and Riley~\cite{kotsialou2018incentivising}, who designed and studied several delegation procedures in the LD setting, the authors studied the ability of their procedures to incentivize participation in the delegation/voting process.

Lastly, in a different direction, several articles have investigated the stability of the delegation process in LD. Indeed, because voters’ preferences over delegates may be conflicting, transitive delegations may lead to an unstable situation as some voters would prefer to revoke their delegations. This question was first studied by Bloembergen, Grossi and Lackner~\cite{bloembergen2019rational} and Escoffier, Gilbert and Pass-Lanneau~\cite{escoffier2018convergence,EscoffierGP20} by using standard concepts from game theory. Kavitha et al.~\cite{kavitha2020popular} studied this problem but using popularity as the stability criterion. Recently, Zhang and Grossi~\cite{zhang2020power} adopted a cooperative game viewpoint of the delegation process in liquid democracy. They defined \textit{delegative simple games} and provided a formal way to measure the importance of both voters and delegators in a delegation graph. Last, they studied delegation games in which the utility that each voter receives depends in part on this measurement. %Then they defined new delegation games, where in a delegation graph the utility that each voter receives depends on this measurement and the accuracy of her guru. They showed that there is a Nash equilibrium in a delegation game when the graph is complete, and however, in general graphs there are some delegation games without Nash equilibrium.

\section{Preliminaries}

We consider a binary election involving a set $\{T,F\}$ of two alternatives and a set $V = \{1,\ldots, n\}$ of $n$ voters. For the sake of simplicity, we assume that the number of voters $n$ is odd. The alternative $T$ denotes the ground truth, i.e., $T$ is more desirable than $F$ and should be elected. 
%\textit{correct } \textit{T} (referred to as the \textit{ground truth}) and one \textit{incorrect} \textit{F}. 
%The set of Let $  V= \left\lbrace 1, 2, ..., n \right\rbrace$ be a finite set of voters that take part in a vote. 
Note that we assume that voters vote independently from one another. 
However, voters do not have a direct access to which alternative is the ground truth. Indeed, we consider a simple model in which each voter $v_i$ has a probability $p_i$ of voting for $T$ if she votes directly. The probability value $p_i$ is called the accuracy of voter $i$ and measures her expertise level. 
We denote by $\vec{p}$ the accuracy vector of size $n$ defined by $\vec{p}[i] = p_i$. 
%This model dates back to the work of Condorcet~\cite{de1785essai} and has been recently investigated to study the LD paradigm~\cite{kahng2018liquid,caragianniscontribution}. 
One might expect that values $p_i$ should be greater than or equal to $0.5$. Indeed, even if a voter is ignorant of a topic he cannot do worse than a random choice. However, similarly as Caragiannis and Micha~\cite{caragianniscontribution}, we allow probabilities to be lower than $0.5$ modeling that fact that some voters may suffer from misinformation on a sensitive topic.

Moreover, we assume the voters to be nodes $V$ of a Social Network (SN) modeled as a directed graph $ G=(V, E) $, i.e., each node in the graph corresponds to a voter $i \in V$ and a (directed) edge $(i, j) \in E $ corresponds to a social relation between $i$ and $j$. The set of out-neighbors (resp.\ in-neighbors) of voter $ i $ is denoted by $\mathtt{Nb}_{out}(i)= \left\lbrace j \in V | (i,j) \in E \right\rbrace$ (resp.\ $\mathtt{Nb}_{in}(i)= \left\lbrace j \in V | (j,i) \in E \right\rbrace$). %$Gamma (i)= \left\lbrace j \in N | (i, j) \in E \right\rbrace $. 

%We use the following running example to make the concepts and problems clearer.

Each voter $i$ has two possible choices: either she can vote directly, or she can delegate her vote to one of her neighbors in $\mathtt{Nb}_{out}(i)$\footnote{In this case, we assume she delegates her vote and all votes she has received through delegations to the same voter.}. 
In the first (resp.\ second) case, she is called a \textit{guru} (resp.\  \textit{follower}). These different choices are formalized by a \textit{delegation function}, which is a function $ d: V \rightarrow V $ such that: 
\begin{itemize}
	\item $ d(i)=j$ if voter $i$ delegates to voter $j \in \mathtt{Nb}_{out}(i)$,
	\item $d(i)=i$ if voter $i$ votes directly.
\end{itemize}
A delegation function $d$ implies a \textit{delegation graph} $H_d$ which is the subgraph of $G$ where there is an edge $(i,j)$ iff $d(i) = j$. Hence, the set of nodes with outdegree $0$ denoted by $Gu(d)$ are the voters who vote directly (i.e., gurus). %equal to the number of her followers plus one (accounting for herself). The election outcome is computed under plurality rule. Delegations partition a graph (SN) into some directed trees (referred to as a \textit{delegation graph}) that are rooted at some gurus (if you reverse the delegation arcs). 
\begin{figure}[t]
\centering
\scalebox{0.8}{\begin{tikzpicture}
\tikzset{vertex/.style = {shape=circle,draw = black,thick,fill = white}}
\tikzset{edge/.style = {->,> = latex'}}

\node[vertex][label=right:{0.9}] (1) at  (0.5,0) {$1$};
\node[vertex][label=left:{0.65}] (2) at  (1.5,-1) {$2$};
\node[vertex][label=above:{0.45}] (3) at  (1.5,1) {$3$};
\node[vertex][label=left:{1}] (4) at  (.5,1.5) {$4$};
\node[vertex][label=right:{0.5}] (5) at (3,1.5) {$5$};
\node[vertex][label=right:{0.35}] (6) at (3.1,0) {$6$};
\node[vertex][label=left:{0.8}] (7) at (3.5,-1) {$7$};
%edges
\draw[edge] (2) to (1);
\draw[edge, <->] (2) to (3);
\draw[edge, <->] (4) to (3);
\draw[edge] (3) to (5);
\draw[edge, <->] (6) to (5);
\draw[edge] (6) to (7);
\draw[edge, <->] (6) to (2);

\node[vertex][label=right:{0.9}, dotted] (B1) at  (5.5,0) {$1$};
\node[vertex][label=left:{}] (B2) at  (6.5,-1) {$2$};
\node[vertex][label=above:{}] (B3) at  (6.5,1) {$3$};
\node[vertex][label=left:{1}, dotted] (B4) at  (5.5,1.5) {$4$};
\node[vertex][label=right:{}]%0.5}, dotted] 
(B5) at (8,1.5) {$5$};
\node[vertex][label=right:{}] (B6) at (8.1,0) {$6$};
\node[vertex][label=left:{0.8}, dotted] (B7) at (8.6,-1) {$7$};
%edges
\draw[edge] (B2) to (B1);
\draw[edge] (B3) to (B4);
\draw[edge] (B5) to (B6);
\draw[edge] (B6) to (B7);
\end{tikzpicture}}
\caption{The SN and voter's accuracy levels (left) and the delegation graph $H_d$ and its gurus dotted (right).} \label{socialpro}
\end{figure}
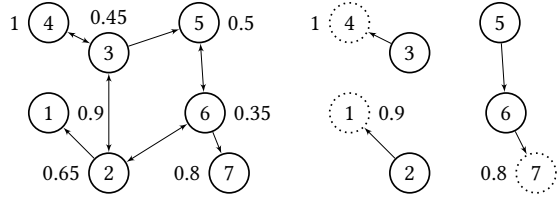
\begin{exmp}\label{probex}
Let us illustrate our notations with an example. We consider an instance with $7$ voters involved in the SN displayed by Figure~\ref{socialpro} and an accuracy vector $\vec{p} = [0.9,0.65,0.45,1,0.5,0.35,0.8]$. 
%The accuracy vector isshows the corresponding SN and the competence levels of the agents which are given next to them.\ Every voter $ i \in \left\lbrace 1, 2, ..., 7 \right\rbrace$ has a competency level $ p_i $, indicating the probability that she will cast a vote for T when she intends to vote directly.

Consider the following delegation function $d$ defined by $d(1)= d(2) = 1$, $d(3)= d(4) = 4$, $d(5) = 6$, and  $d(6)= d(7) = 7$. 
Hence voters $2,3, 5$ and $6$ are followers, and voters $ 1, 4 $ and $ 7 $ are gurus. The graph $H_d$ is represented in Figure~\ref{socialpro}. 
\end{exmp}

We will consider delegation functions $d$ such that $H_d$ is acyclic and we denote by $ \Delta(G)$ the set of \textit{delegation graphs} that can be obtained from $G$. 
Under this assumption, the graph $H_d$ is a forest of directed trees $\{t_1,\ldots,t_{\ell}\}$ such that $t_i$ is rooted at some guru $g_i$. 
Then, each guru $g_i$ receives a voting weight $w(g_i) = |t_i|$, where $|t|$ denotes the number of nodes in tree $t$.
We also extend this notation to subsets $S$ of $Gu(d)$ by letting $w(S) = \sum_{g\in S} w(g)$. 

Given a delegation graph, the election uses a weighted majority rule where guru $g$ has weight $w(g)$. Given $d$ and $\vec{p}$, the probability $P_{d,\vec{p}}[T]$ of electing $ T $ can be computed as follows:
\begin{equation}
   P_{d,\vec{p}}[T] =\sum_{S\subseteq Gu(d)} \prod_{i \in S}p_i \prod_{j \notin S} (1- p_j) \mathbbm{1}_{w(S) > n/2}. \label{eq:definition Pdp} 
\end{equation}
When clear from the context, we will write $P_{d,\vec{p}}$ in place of $P_{d,\vec{p}}[T]$.

In Example~\ref{probex}, as $p_4 = 1$, $T$ wins when voter $1$ or $7$ (or both) votes correctly. Hence, we get that $P_{d,\vec{p}} = 0.98$.

We now make two important remarks.

-- First, note that looking for an optimal delegation function $d$ to maximize $P_{d,\vec{p}}$ is related to looking for an optimal weighting function. By observing that $w(S) > n/2$ if and only if $w(Gu(d) \setminus S) < n/2$, we can conclude that an upper bound to $P_{d,\vec{p}}$ can be obtained if weights are set such that $w(S) > n/2$ when $\prod_{i \in S}p_i \prod_{j \notin S} (1- p_j) > \prod_{j \notin S}p_j \prod_{i \in S} (1- p_i)$. These conditions can be obtained easily if for each guru $g = v_i$, $w(g)$ is proportional to $\log(p_i/(1-p_i))$ \cite{nitzan1982optimal}. Of course, these weights may not be compatible with any delegation function as they may be non-integral or even negative.

-- Second, note that using Equation~\ref{eq:definition Pdp} does not make it possible to compute $P_{d,\vec{p}}$ in polynomial time. However, this can be achieved as follows. 
Start by ordering the gurus in $Gu(d)$ from $g_1$ to $g_{|Gu(d)|}$. 
Then, $P_{d,\vec{p}}$ can be computed by using the following recursive formula where $F(\tau,i)$ denotes the probability that the set of gurus voting for $T$ in $\{g_i,\ldots g_{|Gu(d)|}\}$ has weight at least $\tau$:
\begin{align}\label{lbrecursiveformula}
F(\tau, i) \!=\! \left\{ \begin{array}{c c}
1 & \text{ if }\tau \leq 0 \\
p_{g_{i}}\mathbbm{1}_{w(g_i) \ge \tau} & \text{ if } i=|Gu(d)|\\
p_{g_i}F(\tau -w(g_i),i+1) 
+ & \\ (1-p_{g_i})F(\tau,i+1) & \text{ otherwise.}
\end{array} \right.
\end{align}
Obviously, $P_{d,\vec{p}} = F(\lceil n/2 \rceil,1)$.
To compute $F(\lceil n/2 \rceil,1)$, it suffices to compute values $F(\tau,i)$ for $\tau \in \{0,\ldots, \lceil n/2 \rceil\}$ and $i \in \{1,\ldots, |Gu(d)|\}$ (where $|Gu(d)|\le n$). Hence, using memoization, we get:
% $P_{d,\vec{p}}$ can be computed using $O(n^2)$ operations. 
\begin{proposition}
Given a delegation function $d$, the probability $P_{d,\vec{p}}$ of electing $T$ can be computed using $O(n^2)$ operations. 
\end{proposition}

Following the works of Caragiannis and Micha~\cite{caragianniscontribution} and Kahng et al.~\cite{kahng2018liquid}, we investigate the \textit{Optimal Delegation Problem} (ODP) %where the input is a graph $ G $ and a vector $\vec{p}$, and the 
which aims to coordinate the delegations to maximize $P_{d,\vec{p}}$. 

\vspace{0.2cm}
\begin{center}
\noindent\fbox{\parbox{10cm}{
		\textbf{ODP}\\
		\emph{Input}: A social network $G$ and an accuracy vector $\vec{p}$.\\
		\emph{Feasible Solution}: A delegation function $d$ such that $H_d$ is acyclic. \\
		\emph{Measure}: $P_{d, \vec{p}}$ to maximize.
}}\\

\end{center}
\vspace{0.2cm}

It has been shown by Caragiannis and Micha~\cite{caragianniscontribution} that ODP is hard to approximate within an additive term of $ 1/16 $. 
In the next section, we provide a complementary approximation hardness result. 
Notably, this hardness result will show that from an approximation viewpoint, the complexity of the problem is sensitive to the connectivity of the network as well as the presence of misinformation (i.e., accuracies below 0.5).

\paragraph{Remark.} The ODP problem suggests that a central authority could select the delegations of the voters. This of course does not seem acceptable. In fact, we assume when studying ODP that the network $G$ is more specific than just a social network. We indeed assume that there is an arc between $i$ and $j$ if $i$ knows of $j$ and agrees to delegate to her. Put another way, we assume that all voters specify a subset of neighbors they would accept to delegate to and ask for a central authority to guide their choice. A similar approach has been taken by Gölz et al.~\cite{golz2018fluid}. We do not make any assumption on how voters choose these subsets.

\section{Hardness of ODP}\label{sechrdodp}

For $r \in (0,1)$, let ODP$_{r}$ be the restriction of problem ODP to instances in which all voters have accuracy greater than $r$.
In this section, we show that for any $r \in (0,0.5)$ and any constant $C>0$, ODP$_{r}$ cannot be approximated within a factor of $\alpha \ge (\ln n)^{-C}$ unless $P = NP$. 
This provides a strong approximation hardness result for ODP whenever some voters suffer from misinformation. 
Interestingly, this result is in strong contrasts to ODP$_{r}$ with $r\ge 0.5$, when the direct voting strategy provides a $1/2$-approximation:

\begin{theorem}\label{ODPth}
	For any $r \in (0,0.5)$, for any constant $C$, there is no polynomial-time algorithm for ODP$_{r}$ that achieves an approximation guarantee of $\alpha \ge (\ln n)^{-C}$, unless $P = NP$. 
\end{theorem}

Our result is obtained through a reduction from minimum set cover which cannot be approximated better than within a factor of $(1 - o(1)) \ln N$ (where $N$ is the number of elements in the minimum set cover instance) unless $P = NP$~\cite{dinur2014analytical}.  
Recall that in the minimum set cover problem, we are given a universe $ U = \{x_1,\ldots, x_N\}$ of $N$ elements, and a collection $ S=\left\lbrace S_1, S_2, ..., S_M \right\rbrace $ of subsets of $ U $. The goal is to find the minimum number of sets from $ S $, denoted by $ OPT_{SC} $, that covers all elements of $ U $. 
Note that $M \le 2^{N}$, and that if $N$ or $M$ is bounded by some constant, then the problem can be solved in polynomial time by a brute force approach. 
This fact is used in the reduction to argue that $N$ and $M$ can be assumed larger than some constant. 
The proof follows from a sequence of Lemmas. 
The proofs of these Lemmas are often deferred to the Appendix and we instead focus on the general structure of the argumentation.

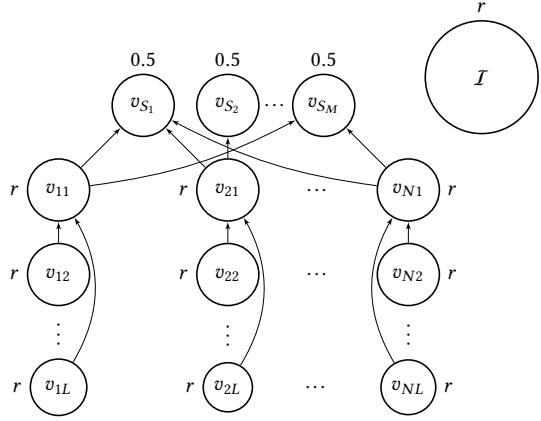
\begin{figure}[t]
\centering
\scalebox{0.75}{
\begin{tikzpicture}
\tikzset{vertex/.style = {shape=circle,draw = black,thick,fill = white,minimum size = 4mm}}
\tikzset{edge/.style = {->,> = latex'}}

\node[vertex][minimum size=1.1cm, label=above:{$0.5$}] (vs1) at  (-.5,0) {$v_{S_1}$};
\node[vertex][minimum size=1.1cm, label=above:{$0.5$}] (vs2) at  (1,0) {$v_{S_2}$};
\node[vertex][minimum size=1.1cm, label=above:{$0.5$}] (vs3) at  (2.7,0) {$v_{S_M}$};

\node[vertex][minimum size=2cm, label=above:{$r$}] (Clique) at (5.5,0.5)  {$\mathcal{I}$};

\node[vertex][minimum size=1.1cm, label=left:{$r$}] (v11) at  (-2,-1.5) {$v_{11}$};
\node[vertex][minimum size=1.1cm, label=left:{$r$}] (v12) at (1,-1.5) {$v_{21}$};
\node[vertex][minimum size=1.1cm, label=right:{$r$}] (v1n) at (4.2,-1.5) {$v_{N1}$};

\node[vertex][minimum size=1.1cm, label=left:{$r$}] (v21) at  (-2,-3) {$v_{12}$};
\node[vertex][minimum size=1.1cm, label=left:{$r$}] (v22) at (1,-3) {$v_{22}$};
\node[vertex][minimum size=1.1cm, label=right:{$r$}] (v2n) at (4.2,-3) {$v_{N2}$};

\node[vertex][minimum size=1cm, label=left:{$r$}] (vm1) at  (-2,-5) {$v_{1L}$};
\node[vertex][minimum size=.5cm, label=left:{$r$}] (vm2) at (1,-5) {$v_{2L}$};
\node[vertex][minimum size=.5cm, label=right:{$r$}] (vmn) at (4.2,-5) {$v_{NL}$};

\path (vs2) to node {\dots} (vs3);
\path (v12) to node {\dots} (v1n);
\path (v22) to node {\dots} (v2n);
\path (vm2) to node {\dots} (vmn);
\path (vm1) -- (v21) node [ midway, sloped] {$\dots$};
\path (vm2) -- (v22) node [ midway, sloped] {$\dots$};
\path (vmn) -- (v2n) node [ midway, sloped] {$\dots$};
\draw[edge] (v11) to (vs1);
\draw[edge] (v11)[bend right = 10] to (vs3);
\draw[edge] (v12) to (vs1);
\draw[edge] (v12) to (vs2);
\draw[edge] (v1n) [bend left = 10] to (vs1);
\draw[edge] (v1n)  to (vs3);

\draw[edge] (v21) to (v11);
\draw[edge] (v22) to (v12);
\draw[edge] (v2n) to (v1n);
\draw[edge] (vm1) [bend right = 30] to (v11);
\draw[edge] (vm2) [bend right = 30] to (v12);
\draw[edge] (vmn) [bend left = 30] to (v1n);
\end{tikzpicture}}
\caption{An example of graph $G$ resulting from the reduction. The set $\mathcal{I}$ includes $K$ isolated nodes with accuracy $r$.} \label{figodphardness}
\end{figure}

\paragraph{The reduction.} 
Let us set $\beta \in (0,0.5)$ and $r = 0.5 - \beta$. 
From an instance $I = (U,S)$ of the minimum set cover problem, we create an instance $I'$ of ODP$_{r}$ as follows. 
As illustrated in Figure~\ref{figodphardness}, the graph $G = (V,E)$ is compounded of the following elements:
\begin{itemize}
    \item{A set $\mathcal{I}$ with $K = 8N^{2}M/\beta^{2}$ isolated nodes, where each node $v \in \mathcal{I}$ has an accuracy of $ r$.}
    \item{$L$ voters $ v_{i1}, v_{i2}, \ldots, v_{iL}$  for each element $ x_i \in U $, where for every $ j \in \{2,\ldots,L\} $, there is directed edge from $v_{ij} $ to $ v_{i1} $ (there are no other edges between these nodes). Each voter $v_{ij} $ has an accuracy of $r$ and the value of $L$ is set to 
    \begin{equation*} 
    L = \Big\lfloor \frac{\beta(4N - 1)}{N(2N-1)}K + \frac{M}{N}\Big\rfloor +1.
    \end{equation*}
	These voters will be called element voters or element nodes in the following.}
	\item{One node $ v_{S_i} $ is created for each $ S_i \in S $ with an accuracy of $ 0.5 $. 
	For every $ i \in [N] $ and $ j \in [M] $, we create a directed edge from $ v_{i1} $ to $ v_{S_j} $ if  $ x_i \in S_j$.
	These voters will be called set voters or set nodes in the following.}
\end{itemize}		 	
		
\textbf{Bound on $|V|$.}  
Let $n = |V|$. Clearly, $n$ is bounded by a polynomial function in $N$ and $M$. 
However, the approximation hardness result on the set cover problem is expressed on $N$ only. 
We thus need to provide a bound on $n$ which only depends on $N$. This bound is provided by the following lemma which is deferred to Appendix~\ref{app : hardness}.
\begin{restatable}{lemma}{boundInN}\label{lemma : boundInN}
If $N \ge 35/\beta^2$, then $n = K +NL + M \le 4^{N}$. 
\end{restatable}

\paragraph{Idea of the reduction.} 
The reduction is built with the following idea: the value of $K$, i.e., the number of nodes in $\mathcal{I}$, is chosen carefully so that with large probability, the ground truth is elected if and only if all element nodes directly or indirectly vote correctly. Indeed, by using Hoeffding inequality (see Lemma~\ref{lem:Hoeffding}), we can give lower and upper bounds on the number of correct votes in $\mathcal{I}$ which are likely to hold. To maximize the probability that all element voters vote correctly, these voters should concentrate their voting power in the hands of as few many set voters as possible, hence looking for a minimum set cover. This connection between the two problems will enable us to show Theorem \ref{ODPth}.

\begin{lemma}(Hoeffding inequality~\cite{hoeffding1994probability})\label{lem:Hoeffding} Let $S$ be the number of successes in $K$ trials of a Bernoulli random variable, which takes value 1 with probability $r$. Then, for every $\epsilon > 0$:
\begin{align*}
Pr[S \ge (r+\epsilon) K ] \le \exp(-2\epsilon^2K),\\
Pr[S \le (r-\epsilon) K ] \le \exp(-2\epsilon^2K). 
\end{align*} 
\end{lemma}

We start formalizing the ideas expressed in the previous paragraph by proving a sequence of lemmas. 
Thereafter,  we set $\epsilon = \frac{\beta}{2(2N-1)}$. 
Lemma~\ref{lem:Hoeffding} shows that it is likely that the number of voters voting correctly in $\mathcal{I}$ belongs to  $((r-\epsilon) K , (r+\epsilon) K)$. 
The next two lemmas, whose proofs are also deferred to Appendix~\ref{app : hardness}, will be used to argue that in that case the ground truth will be elected if all element voters vote (directly or indirectly) correctly. 
\begin{restatable}{lemma}{lemEnough}\label{lem:enough}
If at least $(r - \epsilon)K$ voters in $\mathcal{I}$ vote correctly, then it is enough that all $NL$ element voters vote correctly to elect the ground truth. More formally, we have that
$
(r - \epsilon)K + NL > n / 2
$.
\end{restatable}

\begin{restatable}{lemma}{lemNotEnough}\label{lem:notenough}
Let $K'$ be a number such that $n \ge K' \ge K$. 
If at most $(r + \epsilon)K'$ voters out of $K'$ voters vote correctly, then it is not enough that $\min\{(N-1)L + M, n - K'\}$ other voters vote correctly to elect the ground truth. More formally, if only $(r + \epsilon)K'$ voters out of $K'$ voters vote correctly, then at least $(0.5 - r - \epsilon) K' = (\beta - \epsilon)K'$ voters are already missing and unfortunately, $(N-1)L/2 + M/2 \le (\beta - \epsilon)K'$.
\end{restatable}

We now want to create a connection between a delegation function $d$ in $I'$ and a set cover $X \subset S$ in $I$. 
For this purpose, we introduce a transformation on delegation functions. 
More formally, given a delegation function $d$, we define $\tilde{d}$ as the delegation function obtained from $d$ by making all element voters which are not delegating (directly or indirectly) to a set voter do so\footnote{The choice of  a set voter can be done arbitrarily when several choices are possible.}. 
For a delegation function $d$, let us denote by $X_{d}$ the subset of gurus in $\{v_{S_1}, v_{S_2}, ..., v_{S_M}\}$ that receive some delegations according to $\tilde{d}$.  
Importantly, note that $X_{d}$ corresponds to a set cover in $I$.
  
Let $\mathcal{J}_d$ be the set of voters in $\mathcal{I} \cup \{ v_{i,j} | 1 \le i \le N \text{ and } 2 \le j \le L\}$ which vote directly according to $d$ and $K_{d}  = |\mathcal{J}_d| \ge K$. 
Note that these voters have necessarily a weight of 1 as they may not receive any delegation.
Let $S$ (resp. $S_{d}$) be a random variable representing the number of voters voting correctly in $\mathcal{I}$ (resp. $\mathcal{J}_d$). 
Let $\mathcal{X} =(S \le (r - \epsilon) K )$, 
$\mathcal{Y} = (S \ge (r + \epsilon) K)$,  
and $\mathcal{Y}_d = (S_d \ge (r + \epsilon) K_{d})$. 
Importantly, note that due to lemmas~\ref{lem:enough} and \ref{lem:notenough}, we have that $P_{\tilde{d},\vec{p}}[T | \overline{\mathcal{X}} \cap \overline{\mathcal{Y}}] = 2^{-|X_{d}|}$ (where $\overline{\mathcal{Z}}$ denotes the complement of $\mathcal{Z}$). 
Hence, if $\mathcal{X} \cup \mathcal{Y}$ is a rare event, then $P_{\tilde{d},\vec{p}}$ will be highly dependent on the size of $X_{d}$. 
Interestingly, Lemma~\ref{lem:Hoeffding} (see Appendix~\ref{app : hardness}) allows us to show that events $\mathcal{X} \cup \mathcal{Y}$ and $\mathcal{X}\cup \mathcal{Y}_d$ are indeed rare. 

\begin{restatable}{lemma}{boundRareEvent}\label{lem:bound rare event}
We have the following inequalities:
\begin{align*}
P[\mathcal{X} \cup \mathcal{Y} ] \le  2\exp(-M), \quad \text{ and }\quad
P[\mathcal{X} \cup \mathcal{Y}_d ] \le  2\exp(-M).
\end{align*}
\end{restatable} 

Let $\eta$ be a constant in $(0,1)$. We assume $M$ large enough such that $2\exp(M\ln(2))/(\exp(M) - 2) \le \eta$ and $2\exp(-M) \le \eta$.  
Using Lemma~\ref{lem:bound rare event}, we prove the following relations between $P_{\tilde{d},\vec{p}}$ and $|X_{d}|$.
\begin{restatable}{lemma}{inequalitiesOnPdP}\label{lem:inequalities on pdp}
The following inequalities hold,
\begin{align*}
P_{\tilde{d},\vec{p}} &\ge (1 - 2\exp(-M)) 2^{-|X_{d}|} \ge 2\exp(-M)/\eta,\\ 
P_{\tilde{d},\vec{p}} & \le 3 \times 2^{-|X_{d}|}.
\end{align*}
\end{restatable} 

Lastly, we provide an inequality between $P_{d,\vec{p}}$ and $P_{\tilde{d},\vec{p}}$:
\begin{restatable}{lemma}{inequalitiesOnPdpTwo} \label{lem:inequalities on pdp 2}
    The following inequality holds between $P_{d,\vec{p}}$ and $P_{\tilde{d},\vec{p}}$.
$$P_{d,\vec{p}} \le (1+\eta) P_{\tilde{d},\vec{p}}.$$
\end{restatable}

We are now ready to prove Theorem~\ref{ODPth}.
\begin{proof}[Proof of Theorem~\ref{ODPth}]
Let $OPT_{SC}$ and $P_{d^{*}, \vec{p}}$ be the optimal values in $I$ and $I'$ respectively. 
Moreover, Let $d_{SC}$ be a strategy such that $X_{d_{SC}} = OPT_{SC}$. 
Then  
\begin{align}
P_{d^{*}, \vec{p}} &\ge  P_{\tilde{d}_{SC}, \vec{p}} 
\ge (1 - 2\exp(-M)) 2^{-|X_{d_{SC}}|} 
\ge (1 - \eta)\Big(\frac{1}{2}\Big)^{OPT_{SC}}  \label{equation : ineqdOptSC}   
\end{align}
using Lemma~\ref{lem:inequalities on pdp} and the fact that $2\exp(-M) \le \eta$. 

Let us assume that there exists a polynomial-time approximation algorithm $A$ for ODP with approximation factor  $\alpha \ge (\ln n)^{-C}$ for some positive constant $C$. 
In our reduction, we obtain that $\alpha \ge (\ln n)^{-C} \ge (\ln(4)N)^{-C}$.   
Let $c$ be a constant such that $\eta < c < 1$, and $D = \frac{3(1+\eta)}{1 - \eta}$, then we can always assume that $(\ln(4)N)^{-C}/D \ge N^{-(C+\eta)}$, $OPT_{SC}\ge \frac{C+\eta}{(c-\eta)\ln(2)}$ and $\ln(N)\eta \ge 1$, otherwise, $OPT_{SC}$ is bounded by some constant and we can solve $I$ in polynomial time. 
Hence, from $\alpha \ge (\ln n)^{-C}$ we obtain that:
\begin{align*}
\frac{\ln(\alpha/D)}{\ln(0.5)} 
&\le  \frac{C+\eta}{\ln (2)}\ln(N) 
\le OPT_{SC}(c-\eta)\ln(N)\\
& \le OPT_{SC}(c\ln(N)-1)
\end{align*}
		
Let $d$ be the solution returned by algorithm $A$, we deduce that 
\begin{align*}
&(0.5)^{|X_d|} \ge \frac{P_{\tilde{d}\vec{p}}}{3} \ge \frac{P_{d\vec{p}}}{3(1+\eta)}\\ 
&\ge \frac{\alpha}{3(1+\eta)}P_{d^{*}, \vec{p}} \ge \frac{\alpha(1 - \eta)}{3(1+\eta)}\times(0.5)^{OPT_{SC}}    
\end{align*}
using Lemmas~\ref{lem:inequalities on pdp}, \ref{lem:inequalities on pdp 2} and Equation~\ref{equation : ineqdOptSC}. 
We conclude that
\begin{align*}
&(0.5)^{|X_d|} \ge \alpha/D (0.5)^{OPT_{SC}} \\
&\Rightarrow |X_d| \le \frac{\ln (\alpha/D)}{\ln(0.5)} + OPT_{SC} \le c\ln(N)OPT_{SC}.
\end{align*}
Hence, $A$ would provide a $c \ln(N)$ approximation with $c < 1$ for minimum set cover which is not possible unless $P = NP$. 
\end{proof}

Note that the reduction that we have used creates instances of ODP which have specific properties.  
First, they need to include voters whose accuracy is below $0.5$. 
As previously stated, if all voters have an accuracy greater than or equal to $0.5$, then direct voting or any other delegation strategy would yield a $0.5$-approximation. 
Moreover, the instances generated have low connectivity in the sense that many voters cannot be reached by any other voter. 
We now show that this feature of the reduction cannot be completely removed as the problem admits a $0.5$-approximation if the SN is strongly connected. 
Indeed, if the graph is strongly connected, given any voter $v$ it is possible for all voters to delegate to $v$. Interestingly, in this case, we will see that the simple strategy in which all voters delegate to one of the most ``competent'' voter $v^* \in \text{arg}\max\{p_i|i\in V\}$ leads to a 2-approximation algorithm for ODP. We call this strategy the \emph{best guru strategy}. 
\begin{theorem} \label{th: best guru strategy}
	When the SN is strongly connected, the best guru strategy leads to a $1/2$-approximation algorithm for ODP. 
\end{theorem}

As this result is straightforward when $p_{\max} = \max\{p_i|i\in V\}$ is greater than or equal to $0.5$, we focus on the case where $p_{\max} < 0.5$. In the rest of this section, we show that in this case, the best guru strategy is in fact optimal. 
For this purpose, we will require the following lemma.

\begin{lemma}\label{lemma:pmax}
	Given a confidence vector $\vec{p}$, let $\vec{p}_{\max}$ be the vector obtained from $\vec{p}$ by raising all its entries to $p_{\max}$, then for any delegation function $d$, $P_{d,\vec{p}}(T) \le P_{d,\vec{p}_{\max}}(T)$.
\end{lemma}
\begin{proof}
    Let us consider an arbitrary delegation function $d$. 
    Let us assume that a guru $g_i$ has an accuracy $p_{g_i} < p_{\max}$.  
	We can then use the recursive formula presented in Equation~\ref{lbrecursiveformula} by assuming that we have ordered voters such that voter $g_i$ is first in the ordering. 
	\begin{align*}
	F(\tau >& 0,i)\!=\! p_{g_i} F(\tau \!-\! w(g_i),i\!+\!1) \!+\! (1\!-\!p_{g_i})F(\tau,i\!+\!1)\\
=& F(\tau,i+1) \!+\! p_{g_i} (F(\tau \!-\! w(g_i),i\!+\!1) \!-\! F(\tau,i\!+\!1)).
	\end{align*}
	As $F(\tau',j) - F(\tau,j)\ge 0$, when $\tau'\le \tau$, we conclude that increasing $p_{g_i}$ to $p_{\max}$ can only increase $P_{d,\vec{p}}(T)$. The proof follows by recursion of this argument. 
	%The result then follows from the fact that $F(\tau',j) - F(\tau,j)\ge 0$, when $\tau'\le \tau$.
\end{proof}

Once all the entries of $\vec{p}$ have been raised to $p_{\max}$, we can use a result by Berend and Chernyavsky (Theorem 3 in \cite{berend2008effectiveness}). %to prove that if $p_{\max} \le 0.5$, then ``the best guru strategy'' is optimal. 
This result states that the expert rule (where one voter has all the voting power) is the less effective rule to elect the ground truth when $\vec{p} \ge 0.5$. 
We equivalently use it to state that it is the most effective rule to elect the ground truth when $\vec{p} \le 0.5$. 
Moreover, note that in their setting, an important difference is that weights are not attached to voters but rather distributed uniformly at random before voting. 
However, note that their setup is equivalent to ours when all voters have the same accuracy which is the case when using $\vec{p}_{\max}$. 
Indeed, in this case, the way in which the weights are allocated to voters do not impact $P_{d, \vec{p}}$. 
We may then state the following lemma, which provides the proof for Theorem~\ref{th: best guru strategy}. 
\begin{lemma}
When the SN is strongly connected, the best guru strategy is an optimal solution for ODP when $p_{\max} \!=\! \max\{p_i|i\in V\} \!\le\! 0.5$.% is less than 0.5. 
\end{lemma}

The conclusion of this section is that organizing delegations in an LD framework to maximize the probability of finding the ground truth with provable approximation guarantees is a hard problem. Educating the members of the SN and making them more connected are two levers to address this problem as they can lead to easy instances from an approximation viewpoint. In the next two sections, we will discuss exact and heuristic approaches for ODP.

\section{Exact and Heurisctic Methods} \label{section : heuristics}
\subsection{A MILP for ODP}
The optimal delegation problem can be solved by using the following MILP. 
Let us order the voters such that $V = \{v_1, \ldots, v_n\}$. 
We introduce one binary variable $\delta_{i,w}$ for each $i\in \{1,\ldots, n\}$ and each $w \in \{0,\ldots, n\}$ such that  $\delta_{i,w} = 1$ iff voter $v_i$ receives a weight worth $w$. 
We now introduce new variables and constraints to ensure that these weights are valid and consistent with some delegation function. 
Let variable $z_{i,j,w}$ for $j \in \mathtt{Nb}_{out}(i)$ and $1 \le w \le n$ be a binary variable which equals one iff $i$ delegates $w$ votes to $j$. 
Then, we add the following constraints:
\begin{align}
    \sum_{j \in \mathtt{Nb}_{out}(i)} \sum_{w = 1}^{n} z_{i,j,w}w + \!\! \sum_{w = 0}^{n} \delta_{i,w} w & \notag\\
    - \sum_{k \in \mathtt{Nb}_{in}(i)} \sum_{w = 1}^{n} z_{k,i,w}w &= 1, \forall i \in \{1,\ldots, n\} \label{constraintdelta}\\
    \sum_{j \in \mathtt{Nb}_{out}(i)} \sum_{w = 1}^{n} z_{i,j,w} &= \delta_{i,0}, \forall i \in \{1,\ldots, n\}\\
    \sum_{w = 0}^{n} \delta_{i,w} &= 1, \forall i \in \{1,\ldots, n\}\\
    \sum_{i = 1}^{n} \sum_{w = 0}^{n} \delta_{i,w}w &= n\\
    \delta_{i,w} \in \{0,1\}, \forall (i,w), \quad 
    &z_{i,j,w}  \in \{0,1\}, \forall (i,j,w)
\end{align}
The first constraint is a flow constraint ensuring that the voting power of a voter is equal to the voting power that she receives from other voters plus one. The second constraint ensures that each voter makes only one choice (voting or delegation) and uses all her voting weight accordingly. The third constraint ensures that each voter is allocated some voting weight between $0$ and $n$ and the fourth constraint ensures that all voting weight is used in the voting process, also preventing delegation cycles.  

Then, we introduce $n\times \lceil n/2\rceil$ continuous variables $x_{i,\tau}$ representing the probability that the weights of correct voters in $\{v_i,\ldots, v_n\}$ exceeds threshold $\tau$. 
Of course, our objective is to maximize $x_{1,\lceil n/2 \rceil}$. 
For ease of notation, we introduce some meta-variables $y_{i,\tau}$ where $y_{i,\tau} = 1$ if $\tau \le 0$, $y_{i,\tau} = 0$ if $i > n$ and $\tau > 0$, and $y_{i,\tau} = x_{i,\tau}$ otherwise. 
The variables $y_{i,\tau}$ are linked by the following equations which are reminiscent of the recursive formula used in Equation~\ref{lbrecursiveformula}. 
\begin{align} \label{constraintsy}
    y_{i,\tau} & \le p_i y_{i+1,\tau - w} + (1-p_i)y_{i+1,\tau} + (1 - \delta_{i,w}), \forall (i,\tau,w)
\end{align}
Note that, due to the values taken by $y_{i,\tau}$ when $\tau  < 0$ or $i > n$, the constraint is only binding if $\delta_{i,w}=1$. 
The complete MILP is obtained by maximizing $x_{1,\lceil n/2 \rceil}$ under constraints \eqref{constraintdelta}--\eqref{constraintsy}.

\begin{comment}
written as follows.
\begin{align*}
    \max x_{1,\lceil n/2 \rceil} \hspace{3cm}&\\
    p_i y_{i+1,\tau - w} + (1-p_i)y_{i+1,\tau} + (1 - \delta_{i,w}) &\ge y_{i,\tau}, \forall (i,\tau,w)\\
    \sum_{j \in \mathtt{Nb}_{out}(i)} \sum_{w = 1}^{n} z_{i,j,w}w + \sum_{w = 0}^{n} \delta_{i,w} w &\\
    - \sum_{k \in \mathtt{Nb}_{in}(i)} \sum_{w = 1}^{n} z_{k,i,w}w &= 1, \forall i \in \{1,\ldots, n\}\\
    \sum_{j \in \mathtt{Nb}_{out}(i)} \sum_{w = 1}^{n} z_{i,j,w}&= \delta_{i,0}, \forall i \in \{1,\ldots, n\}\\
    \sum_{w = 0}^{n} \delta_{i,w} &= 1, \forall i \in \{1,\ldots, n\}\\
    \sum_{i = 1}^{n} \sum_{w = 0}^{n} \delta_{i,w}w &= n\\
    \delta_{i,w} &\in \{0,1\}, \forall (i,w)\\
    z_{i,j,w} & \in \{0,1\}, \forall (i,j,w)
\end{align*}
\end{comment}

%\section{Different Mechanisms for ODP}

\subsection{Heuristic Methods for ODP}\label{DiffMech}
In this subsection, we design some centralized heuristic methods. 
The methods maintain a set $S$ of mandatory gurus and iteratively modify $S$ following either a greedy or a local search strategy. Given $S$, we provide two centralized methods to organize the delegations. 
\begin{itemize}
    \item{The first one, {\tt greedy\_delegation}, works in the following way. It considers each guru $g \in S$ in descending order of accuracy value and allocates to $g$ all remaining non-gurus $v \in V \setminus S$ that can reach $g$ in $G[(V \setminus S)\cup\{g\}]$. Voters that cannot reach any guru in $S$ vote for themselves.}
    \item{The second one, {\tt voronoi\_delegation},  works by making each non-guru $v \in V \setminus S$ delegate to the ``closest'' guru in $S$. To take into account accuracies, the ``distance'' between a voter $v$ and a guru $g$  is defined as the length of the shortest path from $v$ to $g$ divided by the accuracy of $g$. This yields a kind of weighted Voronoi graph structure~\cite{graphvoronoi}. Once again, voters who cannot reach any guru in $S$ vote for themselves.}
\end{itemize}
In what follows, the delegation function $d_S$ corresponding to a set $S$ is obtained either by using the {\tt greedy\_delegation} procedure or the {\tt voronoi\_delegation} procedure.

%intend to propose some delegation mechanisms to find the ground truth with high probability. All of the following mechanisms take as input a graph $G(V, E)$, a vector of confidence level $\vec{p}$ and output a delegation graph $H = (V, E')\in \Delta(G)$.

\paragraph{Greedy Heuristics.} Our greedy heuristics start from the delegation function $d_1 = d$ corresponding to the direct voting strategy, and an empty set $S = \{\}$. At each iteration $i$, we determine the node $v$ whose addition to $S$ provides the largest score (i.e., the largest $P_{d_{S\cup\{v\}},\vec{p}}$ value).  If adding $v$ to $S$ results in a positive increment larger than some small $\epsilon$ value, we set $S$ to $S\cup\{v\}$ and update the delegation function $d_i = d_{S\cup \{v\}}$ accordingly. Otherwise the method returns the current delegation function.

\paragraph{Local Search Heuristics.} Given a starting delegation function $d_1 = d$, we initialize $S$ as $Gu(d)$. Then, at each iteration $i$, we determine the single-node addition or removal operation on $S$ which leads to the largest increment value. If $T$ is the set resulting from $S$ by this optimal modification, then $d_i$ is set to $d(T)$. The local search stops when no add or removal operation can result in a positive increment larger than some small $\epsilon$ value.

\subsection{Decentralized Heuristics}\label{sec : decentralized heuristics}
This subsection is devoted to another method, called {\tt emerging}. While this method can be seen as a decentralized heuristic method, we see it more as a way to simulate what could happen in an LD election without any central authority organizing the delegations. In this case, we make the assumption that voters would delegate to voters more informed than them and that they would do so in a stochastic manner. 

More formally, in {\tt emerging}, we assume that each voter $i$ approves the set $A_i:=\{j \in \mathtt{Nb}_{out}(i): p_j > p_i\} \cup \{i\}$ as possible delegates.\footnote{Caragiannis and Micha~\cite{caragianniscontribution} argue that voters with an accuracy value less than 0.5, and hence driven towards the wrong alternative, would probably approve voters that are in fact less accurate than her. We do not make such assumption here as we believe that for some problems there exist objective signs of expertise that may be acknowledged by all voters even when they suffer from misinformation.} Then, each voter $i$ delegates her vote to a voter $j \in A_i$ with probability $p_j/\sum_{k \in A_i}p_k$ and votes with probability $p_i/\sum_{k \in A_i}p_k$. Note that the resulting delegation graph is necessarily acyclic.

\section{Numerical Tests} \label{section : numerical tests}

\begin{comment}
We test our algorithms on the following different types of randomly generated networks:
\begin{itemize}
    \item random networks~\cite{erdds1959random};
    \item scalefree networks~\cite{barabasi1999emergence};
    \item smallworld networks~\cite{watts1998collective};
\end{itemize}
We also try them on the DBLP coauthor network\footnote{http://www.informatik.uni-trier.de/$\sim$ley/db/}. 
We want to investigate the impact of the different methods on the following measures:
\begin{itemize}
    \item number of gurus and distance to guru.
    \item probability of being correct in a majority election.
    \item average accuracy of gurus.
    \item computation time.
\end{itemize}
The methods to test are the following ones:
\begin{itemize}
    \item Direct democracy;
    \item Greedy cap \cite{kahng2018liquid};
    \item $\epsilon$-Greedy strategy with possible abstention;
    \item The MILP;
    \item some heuristics as the voronoi-ODP method.
\end{itemize}
\end{comment}

In this section, we perform simulations to evaluate the performance of the heuristics presented in Sections~\ref{section : heuristics}. Their performance is confronted to the ones of the {\tt GreedyCap} algorithm by Khang, Mackenzie and Procaccia~\cite{kahng2018liquid} and the direct voting strategy (i.e., direct democracy). Our simulations were executed on a compute server running Ubuntu 16.04.5LTS with 24 Intel(R) Xeon(R) CPU E5-2643 3.40GHz cores and a total of 128 GB RAM. Our algorithms are implemented in python using networkx~\cite{hagberg2008exploring} and our code was executed with python version 3.7.6. We used gurobi version 9.0.2 for solving the MILPs in order to obtain the exact solutions to ODP.

\paragraph{Experimental Setting}
We tested our algorithms on randomly generated networks built using the following different models: the $G_{n,m}$ model~\cite{erdds1959random}, i.e., graphs are chosen uniformly at random from the set of graphs with $n$ nodes and $m$ edges; the Barabási–Albert preferential attachment model~\cite{barabasi1999emergence}; and the Newman–Watts–Strogatz small-world model~\cite{watts1998collective}. These two last models generate scale-free networks and small-world networks respectively and hence have properties that are frequently observed in real-world networks.
%; 
%with size $n$ increasing from $51$ to $262$ in steps of $30$.

To be close to a real world setting, voter's accuracies are generated as a mixture of Gaussians, where there is one Gaussian for experts $\mathcal{N}(0.7,0.1)$ ($10\%$ of the voters), one for misinformed voters $\mathcal{N}(0.3,0.1)$ ($20\% $ of the voters) and one for average voters $\mathcal{N}(0.5,0.1)$ ($70\%$ of the voters). These accuracy values are sampled until they are in the interval $(0,1)$. We suppose that each method (except the MILP) does not have access to the exact accuracy values. Indeed, we suppose that voters or the central authority running the method can only approximate these values. The approximation of an accuracy value $p$ is set to the arithmetic mean of the interval $I = [i{\tt prec}, \min((i+1){\tt prec},1)]$ for which $p\in I$, where ${\tt prec}$ is a parameter indicating the precision with which the accuracies can be approximated. When not specified ${\tt prec}$ is set to $0.1$. 

Errorbars in our plots denote $95\%$-confidence intervals. The measurement points in our plots are averages over 50 experiments, 5 generations of random accuracies on each of 10 random graphs generated according to the respective graph model. For the experiments involving the MILP and for testing the impact of the parameter {\tt prec}, in order to further reduce variance, we generate accuracies 10 times on each of the graphs, thus resulting in these two plots containing means of 100 experiments.

The seven heuristic algorithms that we evaluate are: {\tt greedy\_cap}, the {\tt greedyCap}\footnote{This method uses a parameter $\alpha$ set to 1 and a cap function $C: x\rightarrow 10\log(x)^{1/3}$.} algorithm by Khang, Mackenzie and Procaccia~\cite{kahng2018liquid}; {\tt ls\_gr}, {\tt ls\_vo}, {\tt greedy\_gr} and {\tt greedy\_vo} our local search and greedy strategies using either the {\tt greedy\_delegation} method or the {\tt voronoi\_delegation} method to organize delegations; for these last four methods the parameter $\epsilon$ is set to $0.05$; the {\tt emerging} method and {\tt direct\_demo}.

We evaluate the delegation functions returned by the different methods using the following measures: 
the number of gurus; the average distance from voters to their guru (the length of the shortest path between a voter and her guru in $G$); the average accuracy of voters (i.e., the weighted average of gurus' accuracies where each guru is weighted by the number of voters she represents);  and the probability of being correct in a majority election. 
In the figures below, these measures are denoted by {\tt nbOfGurus}, {\tt avgDistance}, {\tt avgAccuracy} and {\tt score} respectively.
%\begin{itemize}
%	\item  and , denoted by  in figures, respectively.
%	\item , denoted by \textit{score} in figures.
%	\item computation time.
%\end{itemize}

\paragraph{Research questions.}
We investigate the following questions. 
\begin{itemize}
    \item How well do the heuristics perform with respect to direct democracy and with respect to the best possible delegation function obtained by the MILP described in Section~\ref{section : heuristics}?
    \item How much does the parameter ${\tt prec}$ impact these results? Stated differently, how well do we need to evaluate the accuracies of voters to have efficient heuristics? 
    \item How much does the number of arcs $m=|E|$ impact these results? Indeed, the theoretical results obtained in Section~\ref{sechrdodp} suggest that connectivity is a key parameter for ODP.
\end{itemize}

\begin{figure} 
	\centering
	\includegraphics[trim=7mm 15mm 4mm 10mm, clip=true, width=0.6\linewidth]{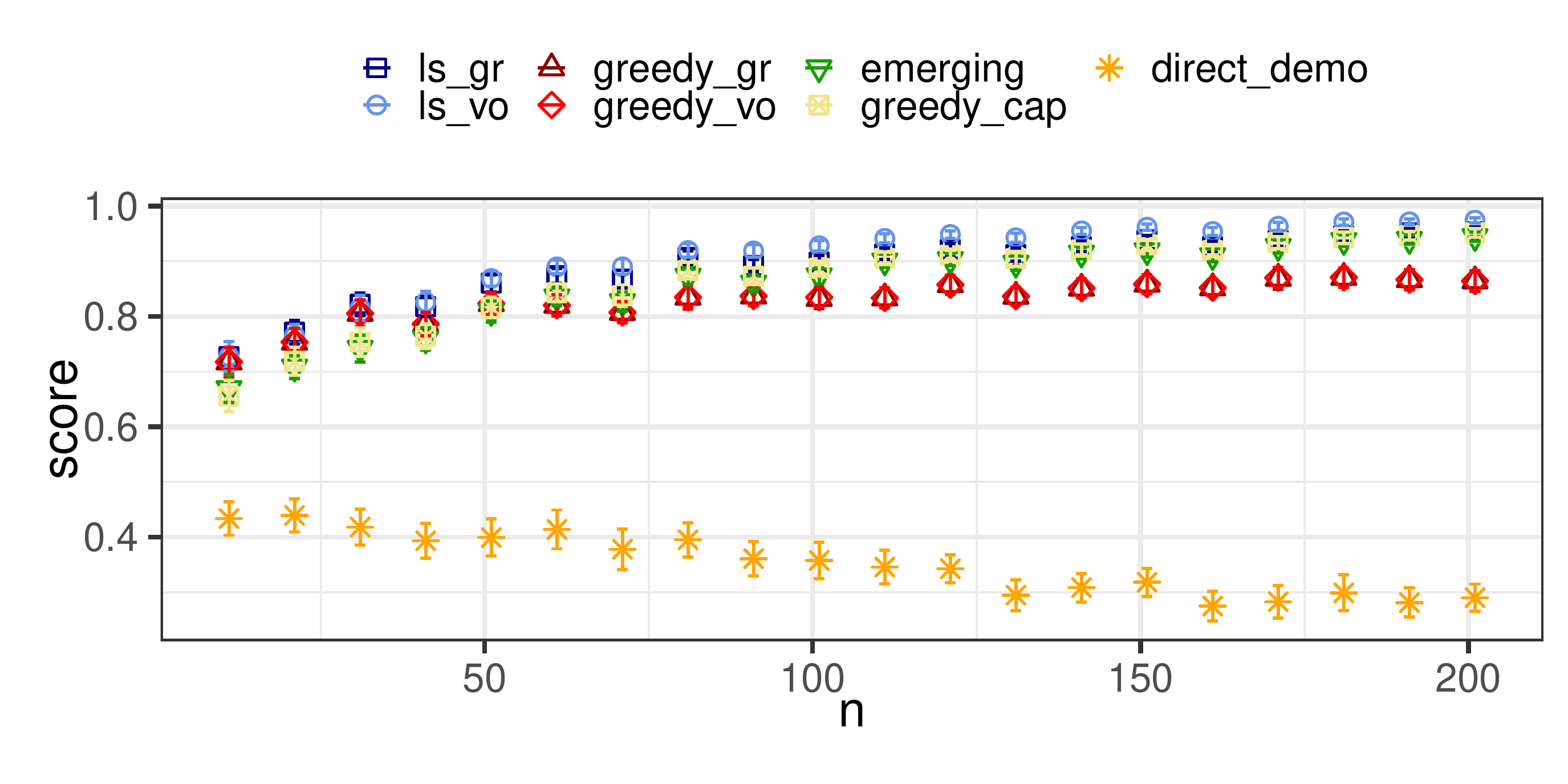}
	\includegraphics[trim=7mm 15mm 4mm 35mm, clip=true, width=0.6\linewidth]{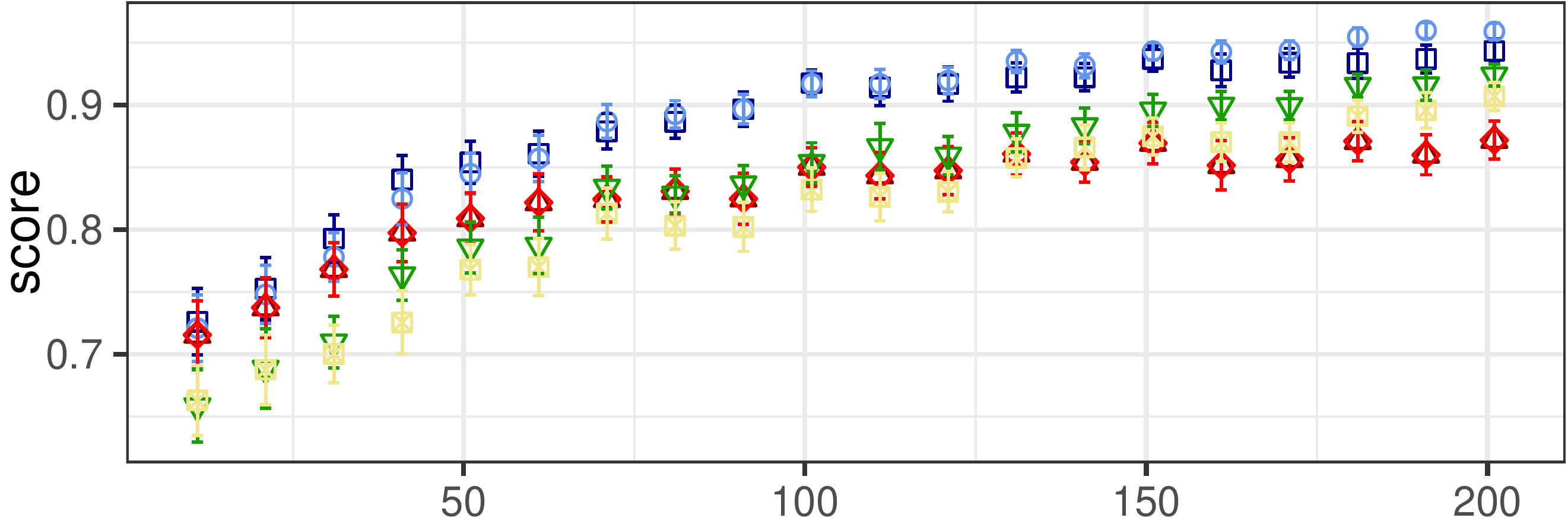}
	\includegraphics[trim=7mm  6mm 4mm 33mm, clip=true, width=0.6\linewidth]{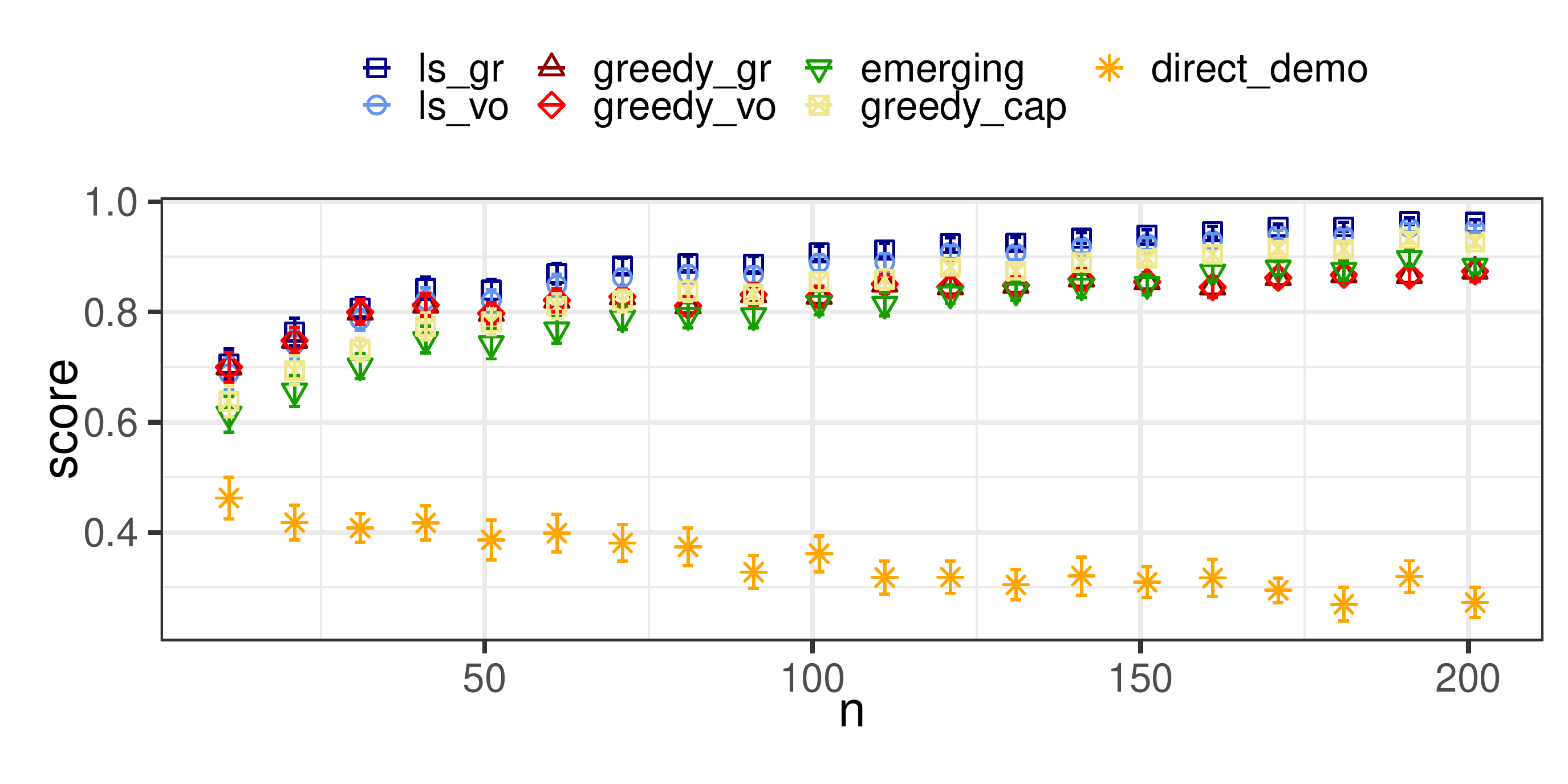}
	\caption{Results for the probability of finding the ground truth using random graphs, $n$ increasing from 11 to 201 in steps of 10 in all three plots: 
	    (1)~$G_{n,m}$ graphs with $m=4n$; 
	    (2)~Barabási–Albert graphs (parameter $m=2$); 
	    (3)~Watts-Strogatz graphs (parameters $k=2$, $p=0.1$).
	}\label{fig : incr_n}
\end{figure}

\paragraph{Results} The probabilities of finding the ground truth resulting from applying our heuristics to random networks with increasing values of $n = |V|$ are plotted in Figure~\ref{fig : incr_n}. We observe that, for all three types of random networks, all heuristics achieve high scores with the local search methods performing best. 
Interestingly, the {\tt greedyCap} and {\tt emerging} methods which are less centralized methods also perform well, electing the ground truth with large probability. In particular, the performance of the {\tt emerging} method suggests that an LD election would lead to a highly accurate decision even without the help of a centralized entity.  
As illustrated in the first plot of Figure~\ref{fig : incr_n}, all these methods outperform by far the direct voting strategy. Indeed, as in our setting, the average accuracy is slightly below $0.5$, the direct voting strategy will perform poorly and its accuracy will not increase in $n$. 
Conversely, as the LD heuristics make it possible to concentrate the voting power in the hands of the most expert voters, we observe that the probability of electing the ground truth increases with the number of such voters and hence in $n$. As this finding also holds in the two other plots, we omit the scores of {\tt direct\_demo} there in order to increase readability. 

To complement these results on the accuracy of our heuristics, we compare the probabilities of finding the ground truth that they yield with the one of the optimal delegation function computed using the MILP presented in Section~\ref{section : heuristics} on small $G_{n,m}$ graphs. In the first plot of Figure~\ref{fig : exact, incr m, incr prec}, we observe that local search strategies seem to provide solutions almost as accurate as the optimal one.    

\begin{figure}[h]
	\centering
	\includegraphics[trim=5mm 8mm 4mm 10mm, clip=true, width=0.6\linewidth]{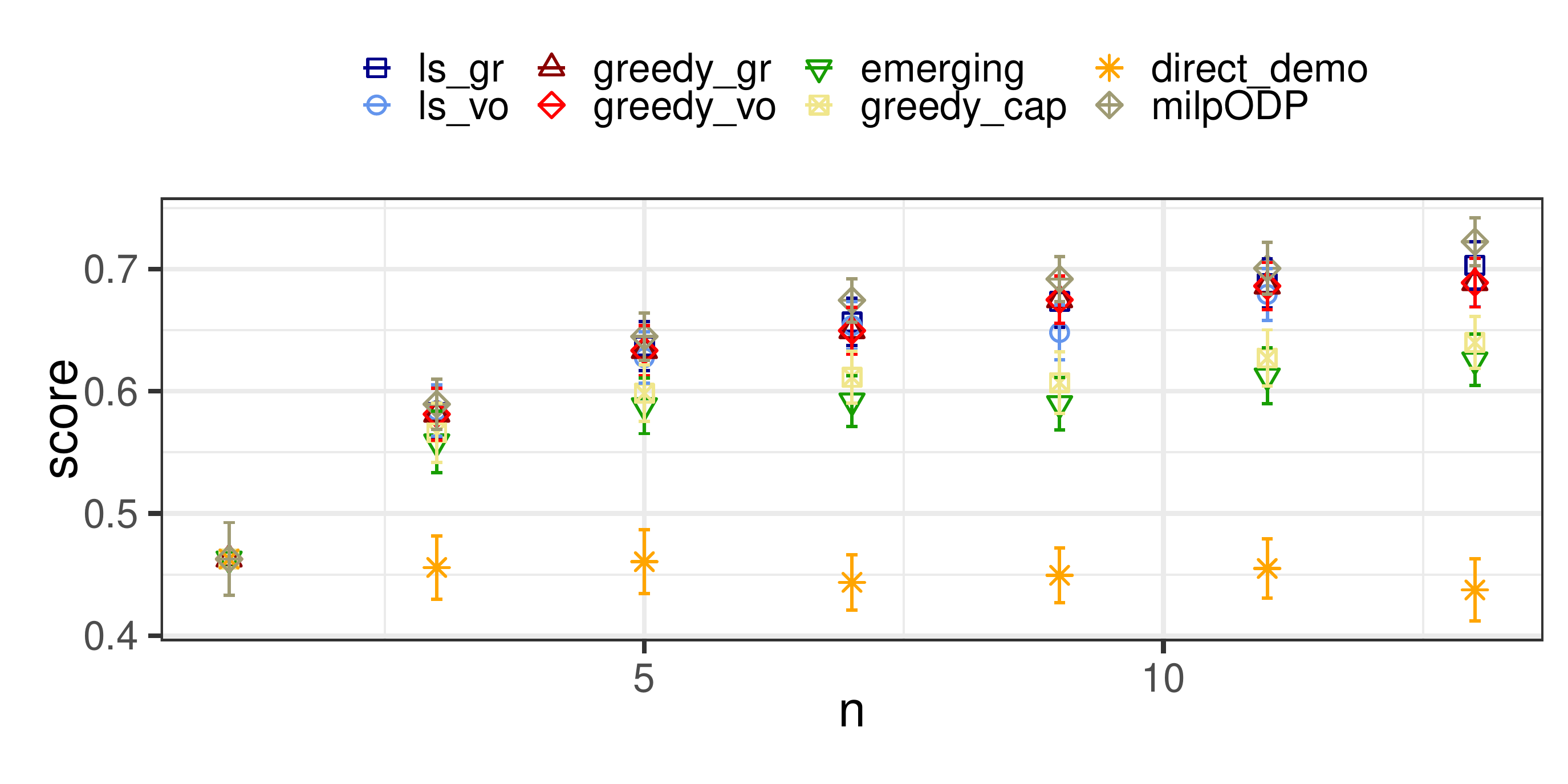}
	\includegraphics[trim=5mm 8mm 2mm 35mm, clip=true, width=0.6\linewidth]{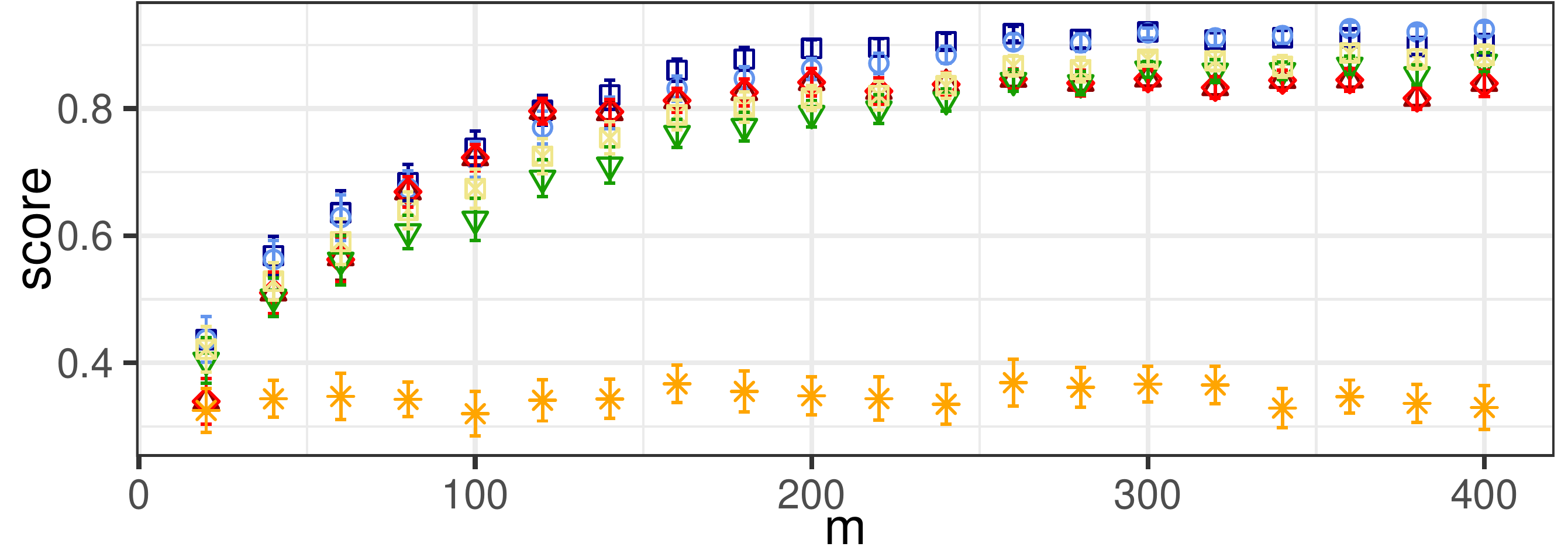}
	\includegraphics[trim=8mm 8mm 2mm 35mm, clip=true, width=0.6\linewidth]{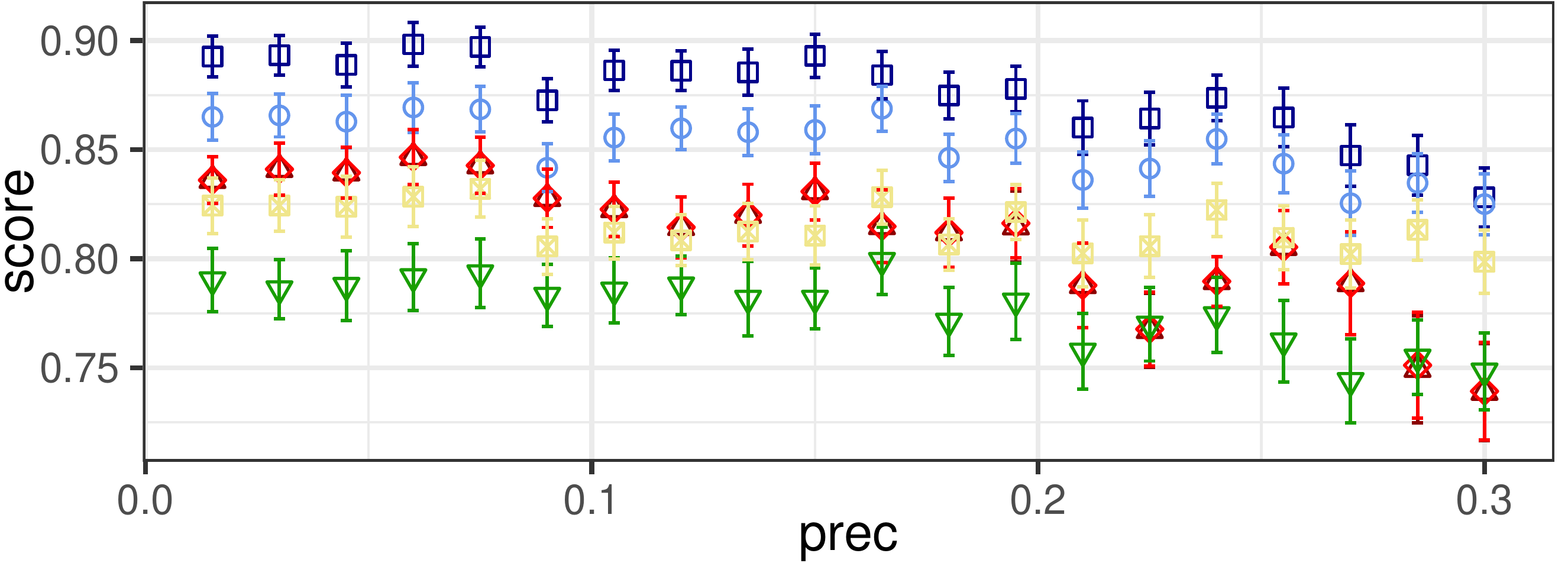}
	\caption{Results for the probability of finding the ground truth using random $G_{n,m}$ graphs with 
	    (1)~$m=2n$ edges, $n$ increasing from 1 to 13 in steps of 2; 
	    (2)~$n=101$ nodes, $m$ increasing from 20 to 400 in steps of 20;
	    (3)~$n=101$ nodes, $m=2n$ and ${\tt prec}$ increasing from $0.015$ to $0.3$ in steps of $0.015$.
	}
	\label{fig : exact, incr m, incr prec}
\end{figure}

Lastly, we evaluate the impact of two other parameters, $m$ and ${\tt prec}$. The probabilities of finding the ground truth resulting from applying our heuristics to $G_{n,m}$ networks with increasing values of $m = |E|$ (resp.\ {\tt prec}) are shown in the second (resp.\ third) plot of Figure~\ref{fig : exact, incr m, incr prec}. On the one hand, we observe that, for all heuristics, their accuracy increases with the connectivity of the network confirming that it is a key feature for \emph{ODP}. Indeed, the more connected the network is, the easier it is for voters to find a suitable guru. On the other hand, increasing {\tt prec} decreases the accuracies of all methods as it becomes increasingly difficult to estimate voters' accuracies. However, even for a large value of {\tt prec} as 0.3, the methods remain quite efficient at finding the ground truth. This suggests that LD can be an efficient collective decision framework, even if voters cannot perfectly evaluate the accuracy of their neighbors.
	
\paragraph{Evaluation on other measures.} In Appendix~\ref{app : numerical tests}, we provide the plots reporting on our evaluations of other measures than {\tt score}. We observe that greedy strategies have much fewer gurus, much larger {\tt avg\_accuracy} values and {\tt avg\_distance} values than other heuristics. Interestingly, all other heuristics yield low {\tt avg\_distance} values and a higher number of gurus which are desirable features for LD's acceptability.

%\subsection{Conclusions on the Evaluation}
%From our simulation, we can conclude that even some simple delegation mechanisms can beat direct democracy outcome by far in average. This supports the important message of LD platform, where we can take the advantage of transferring the power toward appropriate voters in an election in order to have a desire outcome. This message is highlighted more by Emerging-ODP, which is a local mechanism and more natural compared to the other mechanisms. Interestingly and maybe more surprisingly, this message is also highlighted by Genetic-ODP. This mechanism outperforms direct democracy on most of the instances tested and reflects the power of LD platform in leading the power of voters toward suitable gurus.

%\input{contribution/ofdp}
%\input{contribution/k-odp}
%\input{contribution/odpu.tex}
%\input{Observations.tex}

\section{Conclusion}\label{seccln}
In this paper, we have provided new results on the accuracy of the LD paradigm. 
Following a recent work by Caragiannis and Micha~\cite{caragianniscontribution}, we have investigated the \textit{ODP}-problem in which, 
given a binary election with a ground truth, we wish to organize the delegations in a social network in order to maximize the probability 
of electing the correct alternative. 
One the one hand, we have provided a strong approximation hardness result strengthening previously known hardness results for ODP. 
On the other hand we have identified cases in which ODP instances become easy from an approximation viewpoint.
Lastly, we have argued through simulations that simple centralized or decentralized strategies in the LD framework could yield much more accurate decisions than direct democracy.

Several directions of future work are conceivable. First, designing other algorithms that would provide interesting approximation guarantees under some conditions, e.g., on the connectivity of the network, would be a worthwhile contribution. 
Second, it would be interesting to design more evolved decentralized algorithms as we have mostly investigated centralized heuristics in this paper. And yet, it is more natural to expect that most liquid democracy tools will work in a completely decentralized manner. 
Third, it would be interesting to analyze real-world delegation-networks as done in~\cite{kling2015voting} to obtain more insights on the delegation behaviors of voters and how they are related to voters perceptions of their neighbors' expertise level. 
Last, it would be interesting to study the accuracy of the LD framework by using alternative and maybe more complex models than the one of the uncertain dichotomous choice model.
\graphicspath{ {./figures/} }
\bibliography{sample}

\begin{thebibliography}{10}

\bibitem{abramowitz2018flexible}
Ben Abramowitz and Nicholas Mattei.
\newblock Flexible representative democracy: An introduction with binary
  issues.
\newblock In {\em Proceedings of the Twenty-Eighth International Joint
  Conference on Artificial Intelligence, {IJCAI} 2019, Macao, China, August
  10-16, 2019}, pages 3--10, 2019.

\bibitem{barabasi1999emergence}
Albert-L{\'a}szl{\'o} Barab{\'a}si and R{\'e}ka Albert.
\newblock Emergence of scaling in random networks.
\newblock {\em science}, 286(5439):509--512, 1999.

\bibitem{behrens2014principles}
J.~Behrens, A.~Kistner, A.~Nitsche, and B.~Swierczek.
\newblock {\em The principles of LiquidFeedback}.
\newblock Interacktive Demokratie, 2014.

\bibitem{berend2008effectiveness}
Daniel Berend and Yuri Chernyavsky.
\newblock Effectiveness of weighted majority rules with random decision power
  distribution.
\newblock {\em Journal of Public Economic Theory}, 10(3):423--439, 2008.

\bibitem{bloembergen2019rational}
Daan Bloembergen, Davide Grossi, and Martin Lackner.
\newblock On rational delegations in liquid democracy.
\newblock In {\em Proceedings of the AAAI Conference on Artificial
  Intelligence}, volume~33, pages 1796--1803, 2019.

\bibitem{blum2016liquid}
Christian Blum and Christina~Isabel Zuber.
\newblock Liquid democracy: Potentials, problems, and perspectives.
\newblock {\em Journal of Political Philosophy}, 24(2):162--182, 2016.

\bibitem{boldi2009voting}
Paolo Boldi, Francesco Bonchi, Carlos Castillo, and Sebastiano Vigna.
\newblock Voting in social networks.
\newblock In {\em Proceedings of the 18th {ACM} Conference on Information and
  Knowledge Management, {CIKM} 2009, Hong Kong, China, November 2-6, 2009},
  pages 777--786, 2009.

\bibitem{brill2018interactive}
Markus Brill.
\newblock Interactive democracy.
\newblock In {\em Proceedings of the 17th International Conference on
  Autonomous Agents and MultiAgent Systems, {AAMAS} 2018, Stockholm, Sweden,
  July 10-15, 2018}, pages 1183--1187, 2018.

\bibitem{brill2018pairwise}
Markus Brill and Nimrod Talmon.
\newblock Pairwise liquid democracy.
\newblock In {\em Proceedings of the Twenty-Seventh International Joint
  Conference on Artificial Intelligence, {IJCAI} 2018, July 13-19, 2018,
  Stockholm, Sweden.}, pages 137--143, 2018.

\bibitem{caragianniscontribution}
Ioannis Caragiannis and Evi Micha.
\newblock A contribution to the critique of liquid democracy.
\newblock In {\em Proceedings of the Twenty-Eighth International Joint
  Conference on Artificial Intelligence, {IJCAI} 2019, Macao, China, August
  10-16, 2019}, pages 116--122, 2019.

\bibitem{christoffbinary}
Zo{\'{e}} Christoff and Davide Grossi.
\newblock Binary voting with delegable proxy: An analysis of liquid democracy.
\newblock In {\em Proceedings of the 16th Conference on Theoretical Aspects of
  Rationality and Knowledge, {TARK} 2017, Liverpool, UK, 24-26 July, 2017},
  pages 134--150, 2017.

\bibitem{colleysmart}
Rachael Colley, Umberto Grandi, and Arianna Novaro.
\newblock Smart voting.
\newblock In {\em To appear in Proceedings of the Twenty-Nineth International
  Joint Conference on Artificial Intelligence, {IJCAI-PRICAI} 2020, Yokohama,
  Japan, July 11-17,2020}, 2020.

\bibitem{conitzer2013maximum}
Vincent Conitzer.
\newblock The maximum likelihood approach to voting on social networks.
\newblock In {\em 2013 51st Annual Allerton Conference on Communication,
  Control, and Computing (Allerton)}, pages 1482--1487. IEEE, 2013.

\bibitem{de1785essai}
Nicolas De~Condorcet et~al.
\newblock {\em Essai sur l'application de l'analyse {\`a} la probabilit{\'e}
  des d{\'e}cisions rendues {\`a} la pluralit{\'e} des voix}.
\newblock Cambridge University Press, 1785.

\bibitem{dey2021parameterized}
Palash Dey, Arnab Maiti, and Amatya Sharma.
\newblock On parameterized complexity of liquid democracy.
\newblock In Apurva Mudgal and C.~R. Subramanian, editors, {\em Algorithms and
  Discrete Applied Mathematics - 7th International Conference, {CALDAM} 2021,
  Rupnagar, India, February 11-13, 2021, Proceedings}, volume 12601 of {\em
  Lecture Notes in Computer Science}, pages 83--94. Springer, 2021.

\bibitem{dinur2014analytical}
Irit Dinur and David Steurer.
\newblock Analytical approach to parallel repetition.
\newblock In {\em Proceedings of the forty-sixth annual ACM symposium on Theory
  of computing}, pages 624--633, 2014.

\bibitem{erdds1959random}
P~Erd{ö}s and A~R{\'e}nyi.
\newblock On random graphs i.
\newblock {\em Publ. math. debrecen}, 6(290-297):18, 1959.

\bibitem{graphvoronoi}
Martin Erwig.
\newblock The graph voronoi diagram with applications.
\newblock {\em Networks}, 36(3):156--163, 2000.

\bibitem{escoffier2018convergence}
Bruno Escoffier, Hugo Gilbert, and Ad{\`{e}}le Pass{-}Lanneau.
\newblock The convergence of iterative delegations in liquid democracy in a
  social network.
\newblock In {\em Algorithmic Game Theory - 12th International Symposium,
  {SAGT} 2019, Athens, Greece, September 30 - October 3, 2019, Proceedings},
  pages 284--297, 2019.

\bibitem{EscoffierGP20}
Bruno Escoffier, Hugo Gilbert, and Ad{\`{e}}le Pass{-}Lanneau.
\newblock Iterative delegations in liquid democracy with restricted
  preferences.
\newblock In {\em The Thirty-Fourth {AAAI} Conference on Artificial
  Intelligence, {AAAI} 2020, New York, NY, USA, February 7-12, 2020}, pages
  1926--1933, 2020.

\bibitem{estlund1994opinion}
David~M Estlund.
\newblock Opinion leaders, independence, and condorcet's jury theorem.
\newblock {\em Theory and Decision}, 36(2):131--162, 1994.

\bibitem{golz2018fluid}
Paul G{\"{o}}lz, Anson Kahng, Simon Mackenzie, and Ariel~D. Procaccia.
\newblock The fluid mechanics of liquid democracy.
\newblock In {\em Web and Internet Economics - 14th International Conference,
  {WINE} 2018, Oxford, UK, December 15-17, 2018, Proceedings}, pages 188--202,
  2018.

\bibitem{gradstein1986performance}
Mark Gradstein and Shmuel Nitzan.
\newblock Performance evaluation of some special classes of weighted majority
  rules.
\newblock {\em Mathematical Social Sciences}, 12(1):31--46, 1986.

\bibitem{green2005direct}
James Green-Armytage.
\newblock Direct democracy by delegable proxy.
\newblock {\em DOI= http://fc. antioch. edu/\~{} james\_greenarmytage/vm/proxy.
  htm}, 2005.

\bibitem{grofman1983thirteen}
Bernard Grofman, Guillermo Owen, and Scott~L Feld.
\newblock Thirteen theorems in search of the truth.
\newblock {\em Theory and decision}, 15(3):261--278, 1983.

\bibitem{hagberg2008exploring}
Aric Hagberg, Pieter Swart, and Daniel S~Chult.
\newblock Exploring network structure, dynamics, and function using networkx.
\newblock Technical report, Los Alamos National Lab.(LANL), Los Alamos, NM
  (United States), 2008.

\bibitem{hainisch2016civicracy}
Reinhard Hainisch and Alois Paulin.
\newblock Civicracy: Establishing a competent and responsible council of
  representatives based on liquid democracy.
\newblock In {\em 2016 Conference for E-Democracy and Open Government (CeDEM)},
  pages 10--16. IEEE, 2016.

\bibitem{hardt2015google}
S.~Hardt and L.~CR Lopes.
\newblock Google votes: A liquid democracy experiment on a corporate social
  network.
\newblock 2015.

\bibitem{hoeffding1994probability}
Wassily Hoeffding.
\newblock Probability inequalities for sums of bounded random variables.
\newblock In {\em The Collected Works of Wassily Hoeffding}, pages 409--426.
  Springer, 1994.

\bibitem{kahng2018liquid}
Anson Kahng, Simon Mackenzie, and Ariel~D. Procaccia.
\newblock Liquid democracy: An algorithmic perspective.
\newblock In {\em Proceedings of the Thirty-Second {AAAI} Conference on
  Artificial Intelligence, (AAAI-18), New Orleans, Louisiana, USA, February
  2-7, 2018}, pages 1095--1102, 2018.

\bibitem{kavitha2020popular}
Telikepalli Kavitha, Tam{\'a}s Kir{\'a}ly, Jannik Matuschke, Ildik{\'o}
  Schlotter, and Ulrike Schmidt-Kraepelin.
\newblock Popular branchings and their dual certificates.
\newblock In {\em International Conference on Integer Programming and
  Combinatorial Optimization}, pages 223--237. Springer, 2020.

\bibitem{kling2015voting}
Christoph~Carl Kling, J{\'e}r{\^o}me Kunegis, Heinrich Hartmann, Markus
  Strohmaier, and Steffen Staab.
\newblock Voting behaviour and power in online democracy: A study of
  liquidfeedback in germany's pirate party.
\newblock In {\em Proceedings of the 9th International AAAI Conference on Web
  and Social Media(ICWSM).}, pages 208--217, 2015.

\bibitem{kotsialou2018incentivising}
Grammateia Kotsialou and Luke Riley.
\newblock Incentivising participation in liquid democracy with breadth-first
  delegation.
\newblock In {\em Proceedings of the 19th International Conference on
  Autonomous Agents and Multiagent Systems, {AAMAS} '20, Auckland, New Zealand,
  May 9-13, 2020}, pages 638--644, 2020.

\bibitem{nitzan1982optimal}
Shmuel Nitzan and Jacob Paroush.
\newblock Optimal decision rules in uncertain dichotomous choice situations.
\newblock {\em International Economic Review}, pages 289--297, 1982.

\bibitem{paulin2010vzupa}
Alois Paulin.
\newblock {\v{Z}}upa-grassroots e-democracy revolution on the web.
\newblock In {\em International Conference on E-Democracy}, pages 113--123,
  2010.

\bibitem{paulin2020overview}
Alois Paulin.
\newblock An overview of ten years of liquid democracy research.
\newblock In {\em The 21st Annual International Conference on Digital
  Government Research}, pages 116--121, 2020.

\bibitem{shapley1984optimizing}
Lloyd Shapley and Bernard Grofman.
\newblock Optimizing group judgmental accuracy in the presence of
  interdependencies.
\newblock {\em Public Choice}, 43(3):329--343, 1984.

\bibitem{terzopoulou2019optimal}
Zoi Terzopoulou and Ulle Endriss.
\newblock Optimal truth-tracking rules for the aggregation of incomplete
  judgments.
\newblock In {\em International Symposium on Algorithmic Game Theory}, pages
  298--311. Springer, 2019.

\bibitem{watts1998collective}
Duncan~J Watts and Steven~H Strogatz.
\newblock Collective dynamics of ‘small-world’networks.
\newblock {\em nature}, 393(6684):440--442, 1998.

\bibitem{zhang2020power}
Yuzhe Zhang and Davide Grossi.
\newblock Power in liquid democracy.
\newblock {\em arXiv preprint arXiv:2010.07070}, 2020.

\bibitem{zhang2021tracking}
Yuzhe Zhang and Davide Grossi.
\newblock Tracking truth by weighting proxies in liquid democracy.
\newblock {\em arXiv preprint arXiv:2103.09081}, 2021.

\end{thebibliography}

\newpage

\appendix

%\nobalance

\section{Deferred Proofs from Section~\ref{sechrdodp}} \label{app : hardness}
\boundInN*
\begin{proof}
The proof follows from the following inequalities.
\begin{align*}
n & = K + NL + M\\ 
  & = K + N( \lfloor \frac{\beta(4N - 1)}{N(2N-1)}K + \frac{M}{N}\rfloor +1) +M\\
  &\le K + \frac{\beta(4N - 1)}{2N-1}K + N + 2M \quad \text{ as $\lfloor x \rfloor \le x$}\\
  &\le K + 3 K + N + 2M \quad \text{as $\frac{4N - 1}{2N-1} \le 3$ and $\beta<1$}\\
  &\le \frac{(4*8 + 3)N^2M}{\beta^2} \quad \text{ as $N^2M \ge N,M$}\\ 
  &\le \frac{35N^2}{\beta^2}2^N \quad \text{ as $M\le 2^N$}\\
  &\le 4^N \quad \text{as $N\ge 35/\beta^2$ and $N^3\le 2^{N}$ for $N\ge 10$.}  \qedhere
\end{align*} 
\end{proof}

\lemEnough*
\begin{proof}
The result follows from the following inequalities:
\begin{align*}
&(r - \epsilon)K + NL - \frac{n}{2} = (r - \epsilon)K + NL - \frac{K + NL + M}{2}\\
                                   &= -(\beta + \epsilon)K + \frac{NL}{2} - \frac{M}{2}\\
                                   &= - (\beta + \epsilon)K +  \frac{N}{2}(\lfloor \frac{\beta(4N - 1)}{N(2N-1)}K + \frac{M}{N}\rfloor +1) - M/2\\
                                   & > - \beta (1 + \frac{1}{2(2N-1)})K +  \frac{\beta(4N - 1)}{2(2N-1)}K \text{, as $\lfloor x \rfloor + 1 > x$} \\
                                   & = 0.\qedhere
\end{align*}
\end{proof}

\lemNotEnough*
\begin{proof}
The result follows from the following inequalities:
\begin{align*}
&(\beta - \epsilon)K' - \frac{N-1}{2}L - \frac{M}{2} \\
&\ge (\beta - \epsilon)K - \frac{N-1}{2}(\lfloor \frac{\beta(4N - 1)}{N(2N-1)}K + \frac{M}{N}\rfloor +1)  - \frac{M}{2} \\ %\quad \text{ as $K' \ge K$}\\
 & \ge (\beta - \epsilon)K - \frac{N-1}{2}(\frac{\beta(4N - 1)}{N(2N-1)}K + \frac{M}{N} +1) - \frac{M}{2} \\
 & = \beta \frac{4N-3}{2(2N-1)}K - \frac{\beta(4N - 1)(N-1)}{2N(2N-1)}K \\
 & \hspace{3cm} - \frac{M(N-1)}{2N} - \frac{N-1}{2} - \frac{M}{2}\\
 & = \frac{\beta}{2N(2N-1)} ((4N-3)N - (4N - 1)(N-1))K \\
 & \hspace{3cm} - M\frac{2N-1}{2N} - \frac{N-1}{2} \\
 & = \frac{\beta}{2N} K - M\frac{2N-1}{2N} - \frac{N-1}{2} \\
 & \ge \frac{4}{\beta}NM - M\frac{2N-1}{2N} - \frac{N-1}{2} \\ 
 & > 0.\qedhere
 \end{align*}
\end{proof}

\boundRareEvent*
\begin{proof}
We provide the proof for the first inequality. The second one can be proven similarly.
\begin{align*}
P[\mathcal{X} \cup \mathcal{Y}] &\le P[ \mathcal{X}] + P[\mathcal{Y}] \quad \text{ using a union bound}\\ 
&\le 2\exp(-2\epsilon^2K) \quad \text{ using Lemma~\ref{lem:Hoeffding}}\\
&= 2\exp(-\frac{\beta^2}{2(2N-1)^2}\frac{8N^2M}{\beta^2})\\
&\le 2\exp(-M) \qedhere
\end{align*}
\end{proof}

\inequalitiesOnPdP*
\begin{proof}
The proof follows from the following inequalities:
\begin{align*}
P_{\tilde{d},\vec{p}} & = P[\overline{\mathcal{X}} \cap \overline{\mathcal{Y}}] 2^{-|X_{d}|} + P[\mathcal{X} \cup \mathcal{Y}]P_{\tilde{d},\vec{p}}[T | \mathcal{X} \cup \mathcal{Y}] \\
                                & \ge (1 - 2\exp(-M)) 2^{-|X_{d}|}\\
                                & \ge (1 - 2\exp(-M)) 2^{-M} \\
                                & \ge 2\exp(-M)/\eta \quad \text{ as $\frac{2\exp(M\ln(2))}{(\exp(M) - 2)} \le \eta$},
\end{align*}
and,
\begin{align*}
P_{\tilde{d},\vec{p}} & = P[\overline{\mathcal{X}} \cap \overline{\mathcal{Y}}] 2^{-|X_{d}|} + P[\mathcal{X} \cup \mathcal{Y}]P_{\tilde{d},\vec{p}}[T | \mathcal{X} \cup \mathcal{Y}] \\ 
                                & \le 2^{-|X_{d}|} + 2\exp(-M)\\ 
                                & \le 3 \times 2^{-|X_{d}|} \quad \text{ as $e > 2$ and $M\ge |X_{d}|$}. \qedhere
\end{align*}
\end{proof}

\inequalitiesOnPdpTwo*
\begin{proof}
We first want to prove that $P[\overline{\mathcal{X}} \cap \overline{\mathcal{Y}_d} ]P_{d,\vec{p}}[T |\overline{\mathcal{X}} \cap \overline{\mathcal{Y}_d}] \le P[\overline{\mathcal{X}} ]P_{\tilde{d},\vec{p}}[T | \overline{\mathcal{X}}]$. 
First, it is clear that $P[\overline{\mathcal{X}} \cap \overline{\mathcal{Y}_d} ] \le P[\overline{\mathcal{X}} ]$. 
Moreover, by Lemma~\ref{lem:enough}, it is also clear that $P_{\tilde{d},\vec{p}}[T | \overline{\mathcal{X}}] \ge 2^{-|X_d|}$. 
Let $Q_d$ be the set of gurus in $Gu(d) \setminus \mathcal{J}_d$ receiving some delegations from element voters or being an element voter. 
Then, by Lemma~\ref{lem:notenough}, it is clear that $P_{d,\vec{p}}[T |\overline{\mathcal{X}} \cap \overline{\mathcal{Y}_d}] \le 2^{-|Q_d|}$ because if one guru in $Q_d$ does not vote correctly, we are sure that the cumulative weight of correct gurus in $Q_d$ is below $M + (N-1)L$ (this is due to the structure of the social network).  
The claim then follows as $2^{-|Q_d|} \le 2^{-|X_d|}$. 
The proof now follows from the following inequalities:
\begin{align*}
P_{d,\vec{p}} & = P[\overline{\mathcal{X}} \cap \overline{\mathcal{Y}_d}] P_{d,\vec{p}}[T |\overline{\mathcal{X}} \cap \overline{\mathcal{Y}_d}] + P[\mathcal{X} \cup \mathcal{Y}_d] P_{d,\vec{p}}[T |\mathcal{X} \cup \mathcal{Y}_d]\\
                      & \le P[\overline{\mathcal{X}} \cap \overline{\mathcal{Y}_d} ]P_{d,\vec{p}}[T |\overline{\mathcal{X}} \cap \overline{\mathcal{Y}_d}] + P[\mathcal{X} \cup \mathcal{Y}_d]\\
                      & \le P[\overline{\mathcal{X}} ]P_{\tilde{d},\vec{p}}[T | \overline{\mathcal{X}}] + 2\exp(-M)\\ 
                      & \le P_{\tilde{d},\vec{p}} + 2\exp(-M)\\ 
                      & \le (1+\eta) P_{\tilde{d},\vec{p}},
\end{align*}
where we used Lemma~\ref{lem:inequalities on pdp} for the last inequality.
\end{proof}

\section{Deferred Material from Section~\ref{section : numerical tests}} \label{app : numerical tests}
\begin{figure}[p]
	\centering
    \includegraphics[trim=7mm 15mm 4mm 10mm, clip=true, width=.6\linewidth]{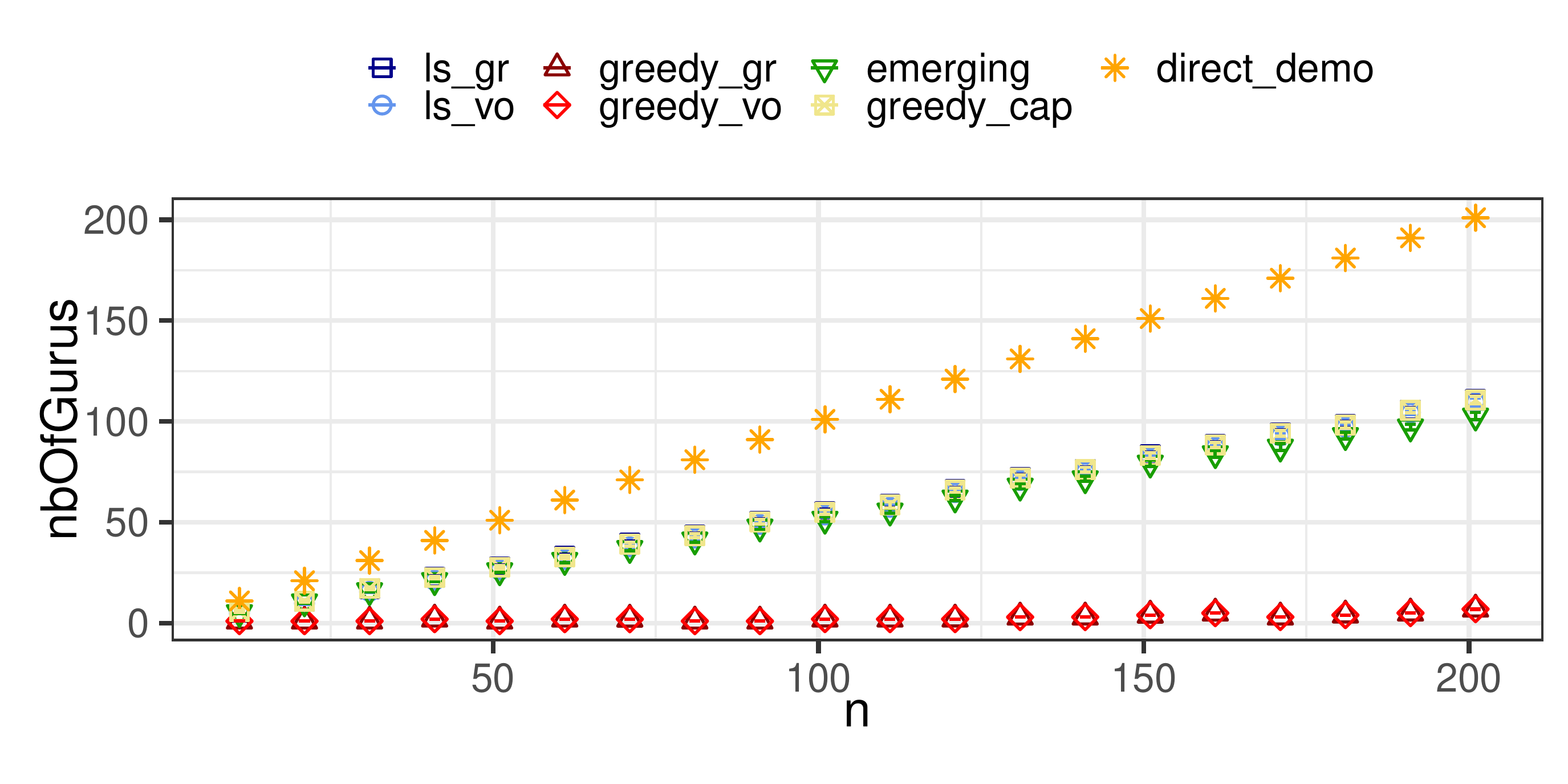}
	\includegraphics[trim=7mm 15mm 4mm 35mm, clip=true, width=.6\linewidth]{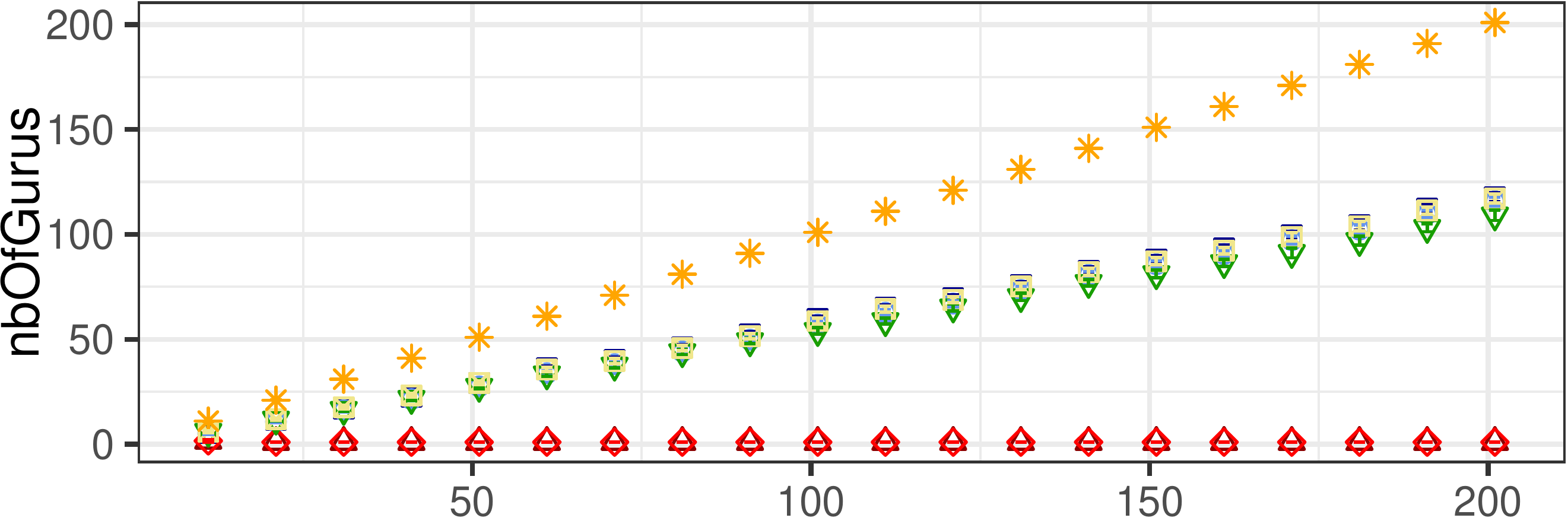}
	\includegraphics[trim=7mm 15mm 4mm 35mm, clip=true, width=.6\linewidth]{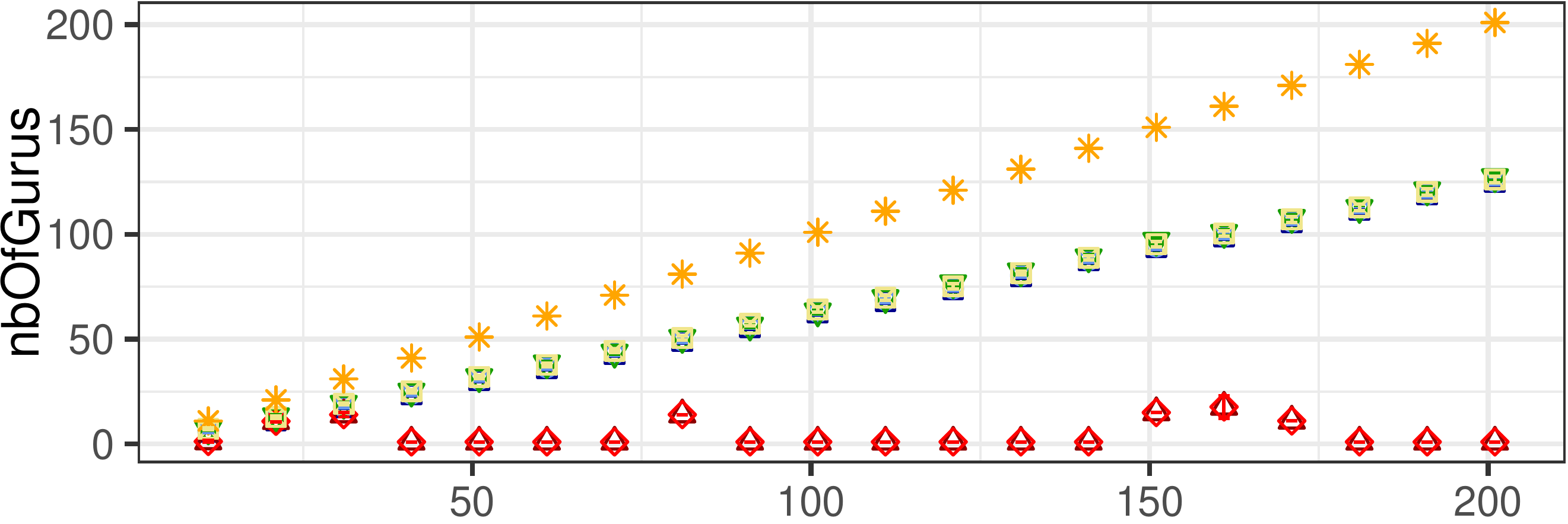}
	\includegraphics[trim=7mm 15mm 4mm 35mm, clip=true, width=.6\linewidth]{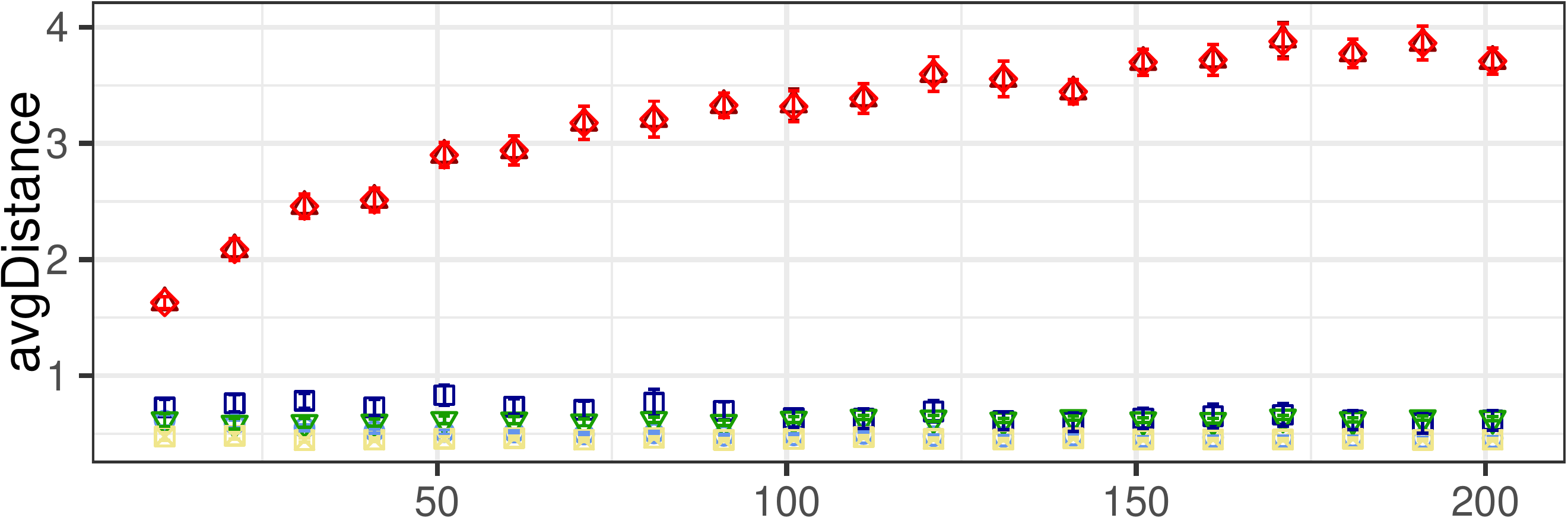}
	\includegraphics[trim=7mm 15mm 4mm 35mm, clip=true, width=.6\linewidth]{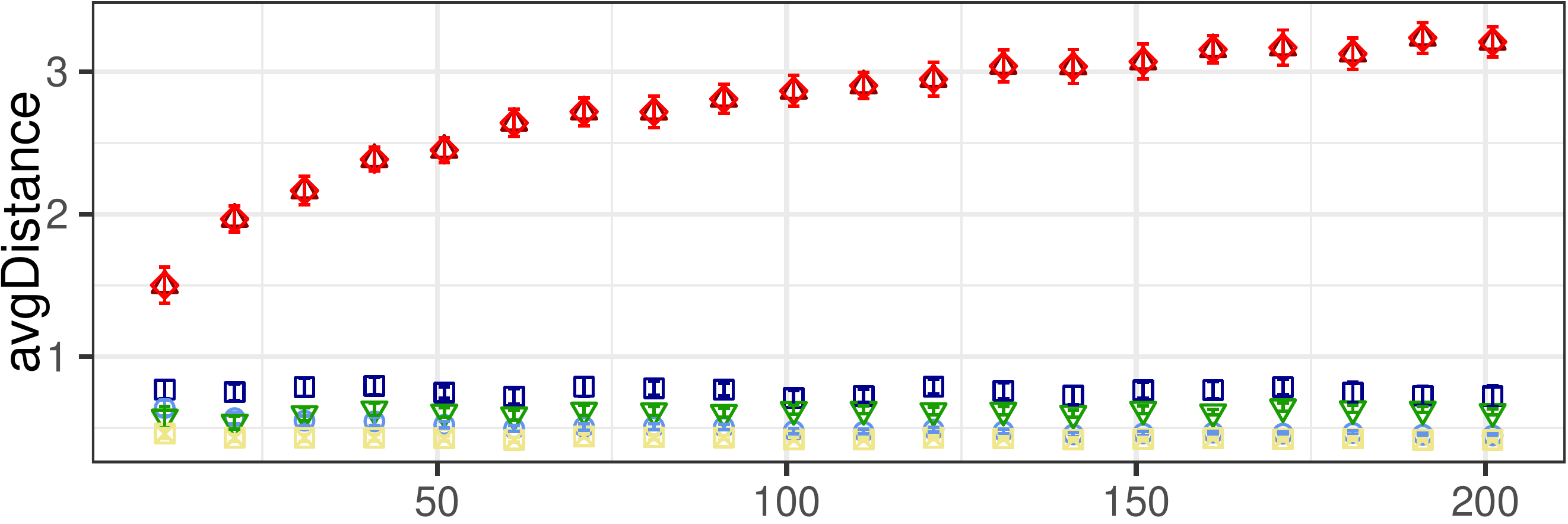}
	\includegraphics[trim=7mm  6mm 4mm 35mm, clip=true, width=.6\linewidth]{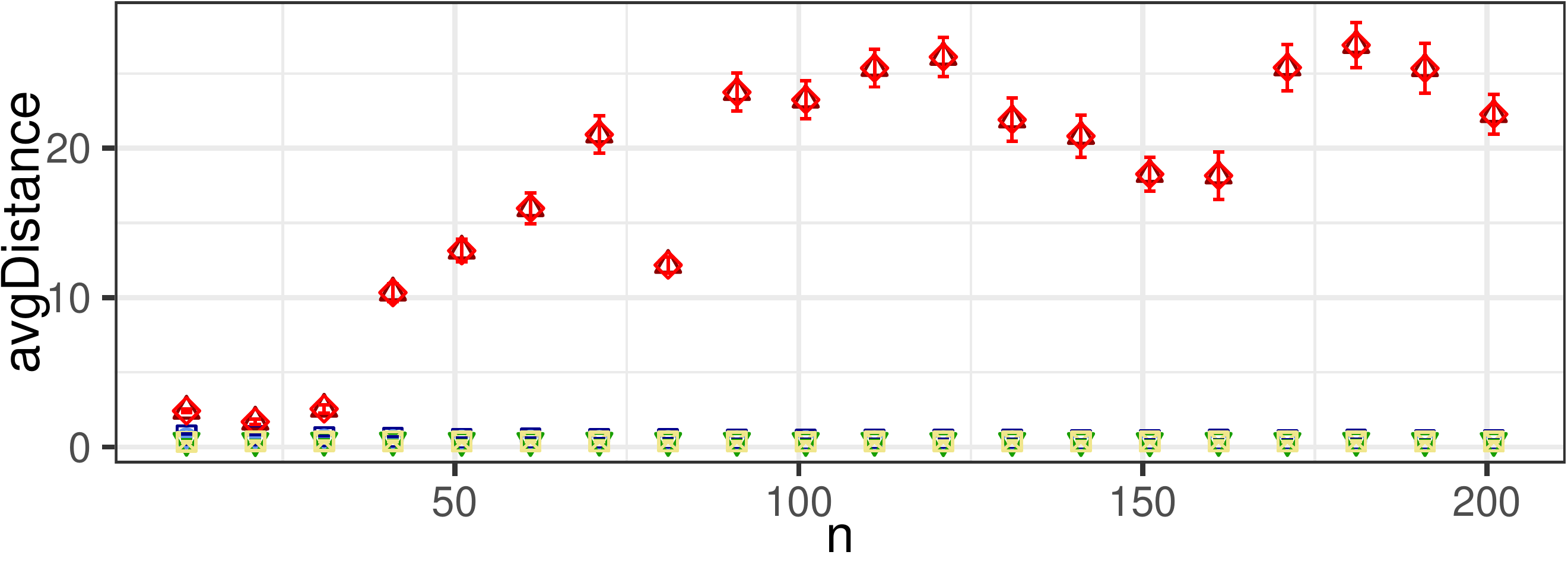}
	\caption{Results for number of gurus and the average distance from voters to their guru using random graphs, $n$ increasing from 11 to 201 in steps of 10 in all six plots: 
	    \mbox{(1, 4)}~$G_{n,m}$ graphs with $m=4n$; 
	    \mbox{(2, 5)}~Barabási–Albert graphs (parameter $m=2$); 
	    \mbox{(3, 6)}~Watts-Strogatz graphs (parameters $k=2$, $p=0.1$).
	} \label{fig : number of gurus average distance n_incr}
\end{figure}

\begin{figure}[p]
	\centering
	\includegraphics[trim=7mm 15mm 4mm 10mm, clip=true, width=.6\linewidth]{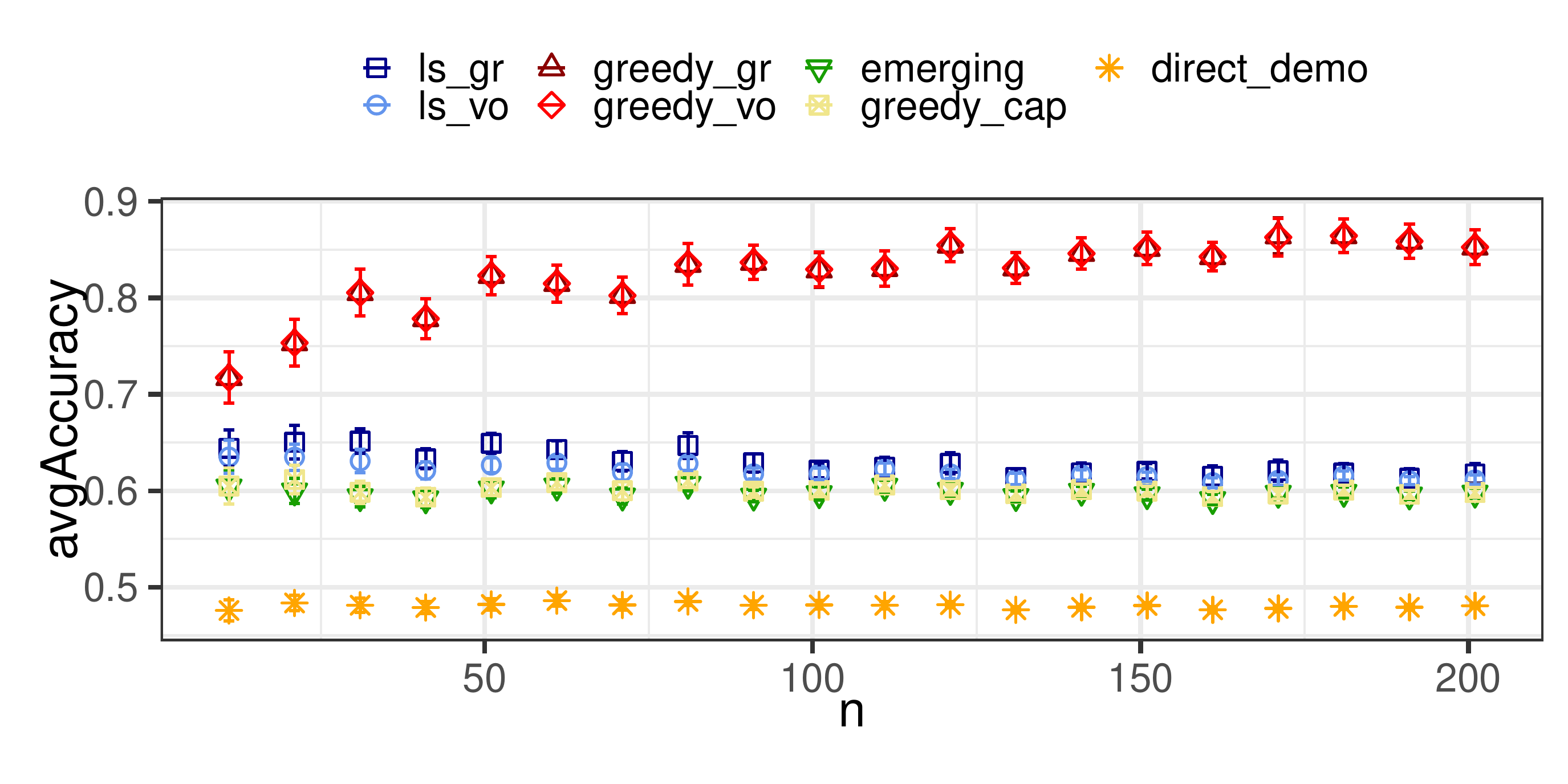}
	\includegraphics[trim=7mm 15mm 4mm 33mm, clip=true, width=.6\linewidth]{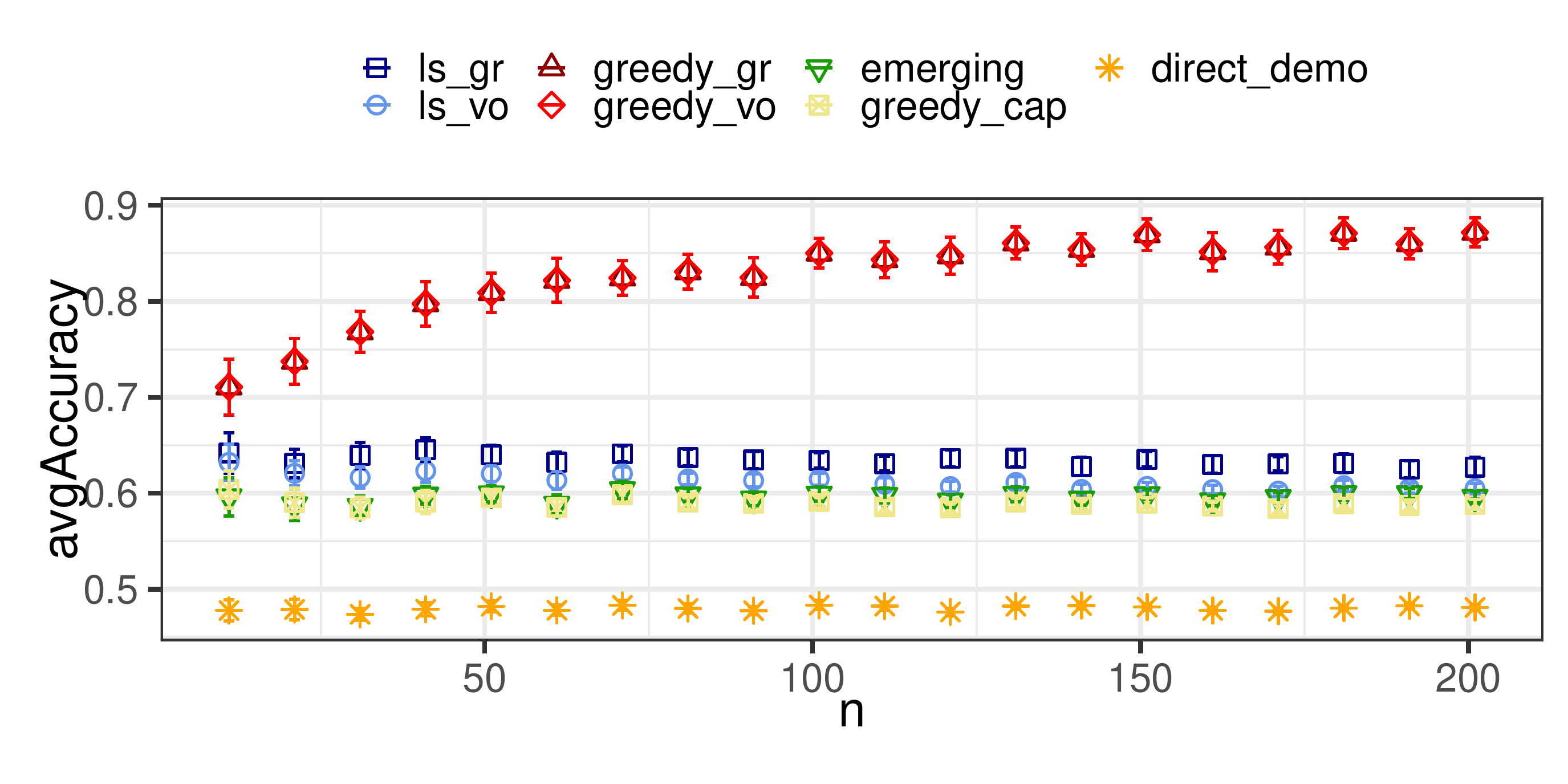}
	\includegraphics[trim=7mm 15mm 4mm 35mm, clip=true, width=.6\linewidth]{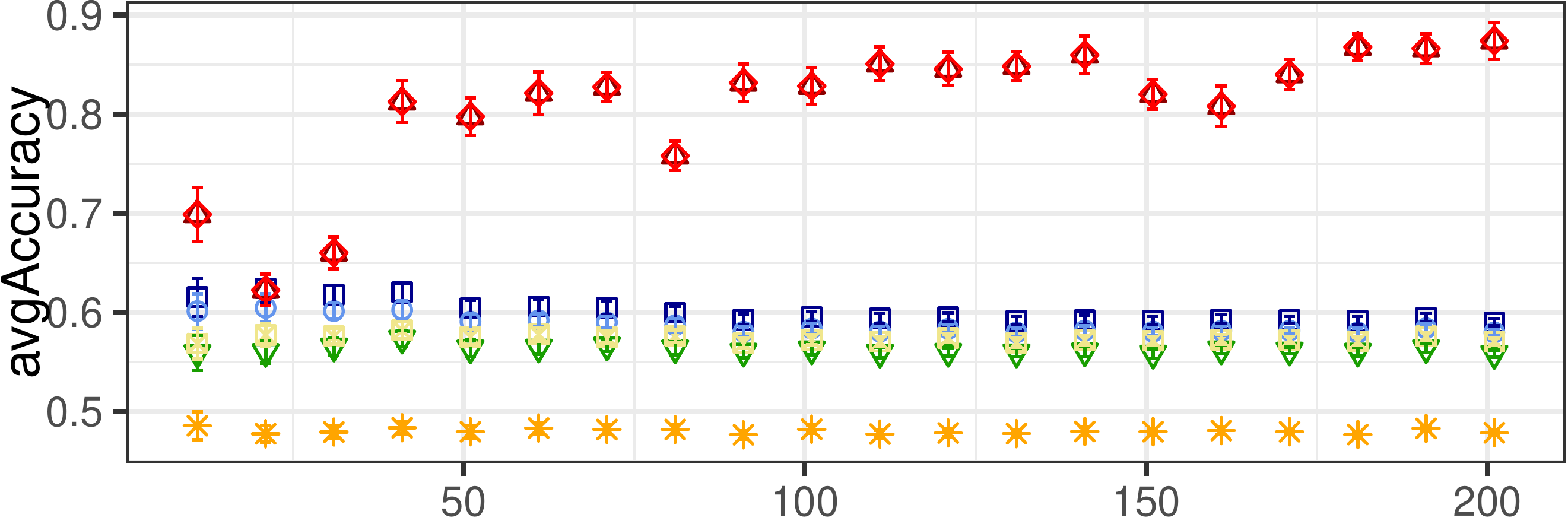}
	\includegraphics[trim=7mm 15mm 4mm 35mm, clip=true, width=.6\linewidth]{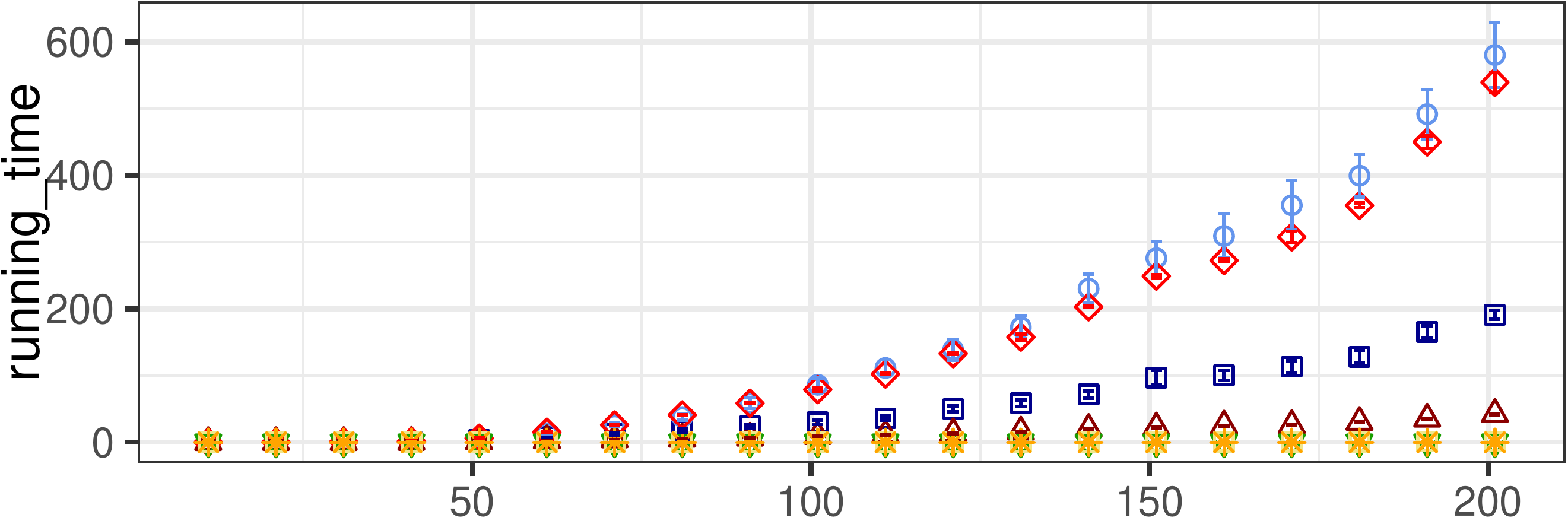}
	\includegraphics[trim=7mm 15mm 4mm 35mm, clip=true, width=.6\linewidth]{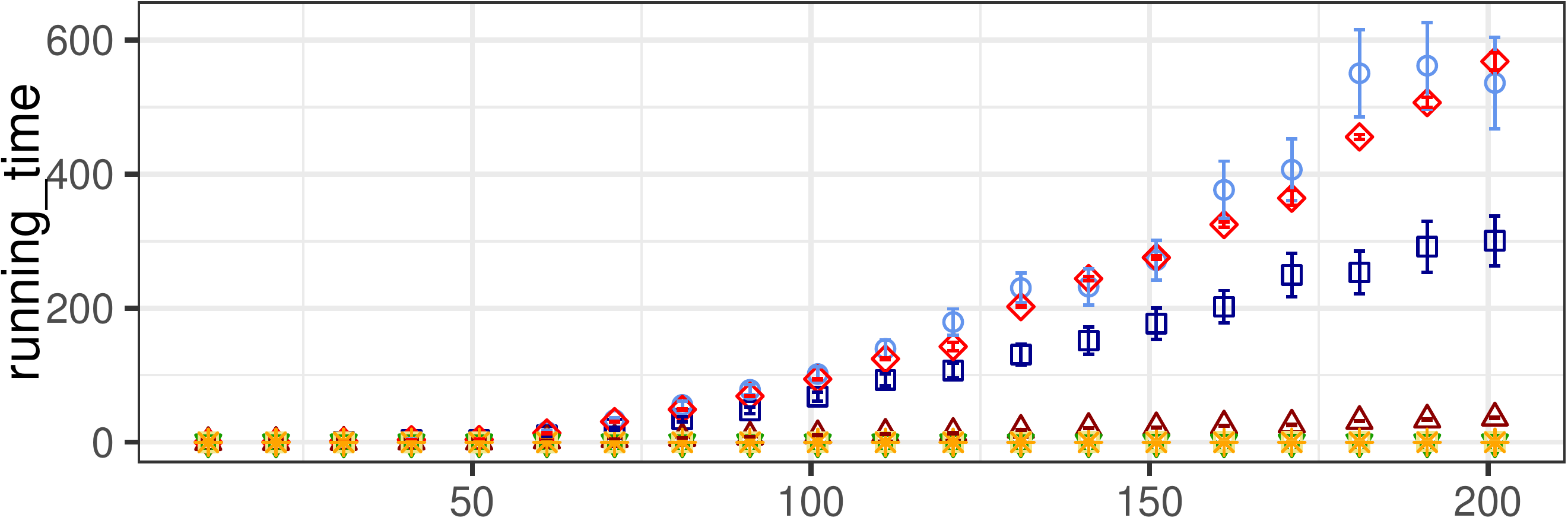}
	\includegraphics[trim=7mm  6mm 4mm 35mm, clip=true, width=.6\linewidth]{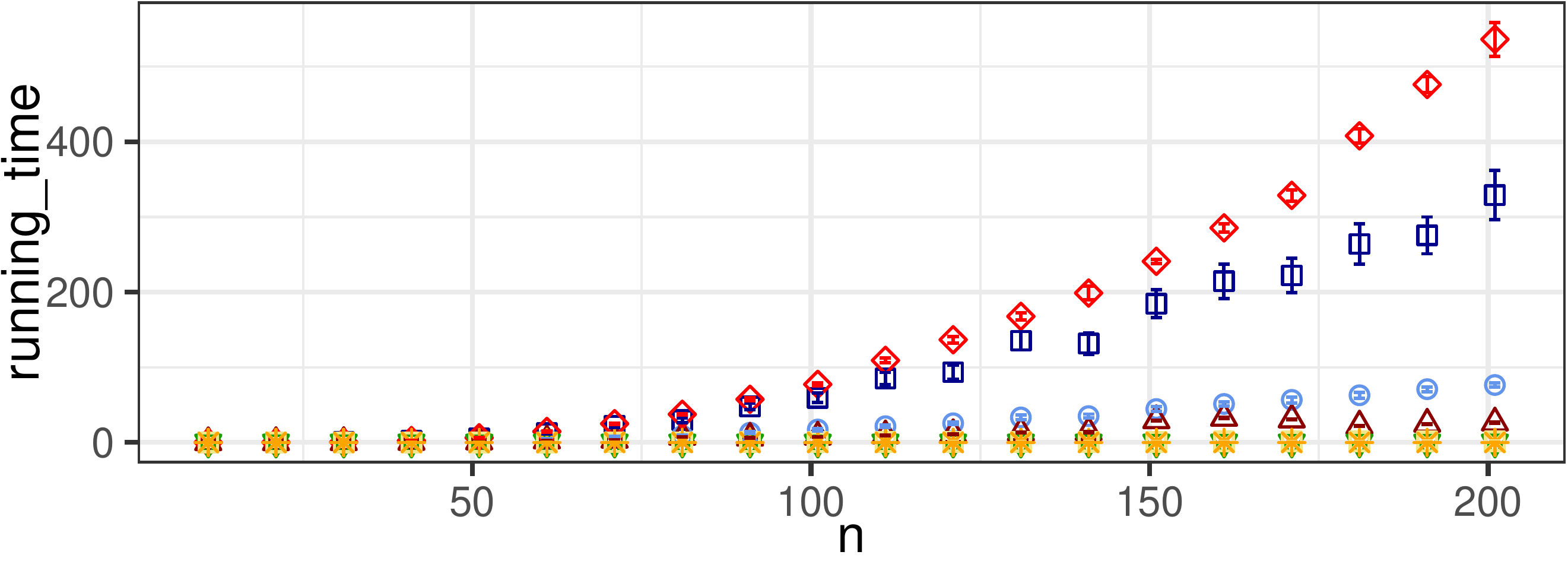}
	\caption{Results for average voter accuracy and running times in seconds of the different methods using random graphs, $n$ increasing from 11 to 201 in steps of 10 in all siz plots: 
	    \mbox{(1, 4)}~$G_{n,m}$ graphs with $m=4n$; 
	    \mbox{(2, 5)}~Barabási–Albert graphs (parameter $m=2$); 
	    \mbox{(3, 6)}~Watts-Strogatz graphs (parameters $k=2$, $p=0.1$).
	} \label{fig : accuracy running time n_incr}
\end{figure}

In this section, we provide the plots reporting on our evaluations of other measures than {\tt score}. As can be seen on Figure~\ref{fig : number of gurus average distance n_incr}~(1,2,3), the greedy strategies select much fewer gurus than other heuristics. This is because they start from an empty set $S$ specifying the mandatory gurus and then get stuck at an early iteration of the algorithm. As a result greedy strategies have much larger {\tt avgDistance} values and {\tt avgAccuracy} values than other heuristics, see Figure~\ref{fig : number of gurus average distance n_incr}~(4,5,6) and Figure~\ref{fig : accuracy running time n_incr}~(1,2,3), respectively. Interestingly, as can be seen in Figure~\ref{fig : number of gurus average distance n_incr}~(4,5,6), all other heuristics yield low {\tt avgDistance} values and a higher number of gurus. These are both desirable features for LD's acceptability. 

As illustrated by Figure~\ref{fig : accuracy running time n_incr}~(1,2,3), our heuristics identify a set of gurus with average accuracy higher than the average accuracy of the population and in particular higher than $0.5$. This is the reason behind the high {\tt score} values that our heuristics achieve. Greedy strategies yield the most expert gurus, but they do not provide very accurate solutions as their number of gurus is not high enough.

Some information on the running times of the methods is provided in Figure~\ref{fig : accuracy running time n_incr}~(4,5,6). We stress that our code is for proof of concept and could benefit from many optimizations to decrease the running times. We observe that local search and greedy strategies using the {\tt voronoi\_delegation} strategy are the most expensive.

\begin{figure}[h]
	\centering
	\includegraphics[trim=7mm 15mm 4mm 10mm, clip=true, width=.6\linewidth]{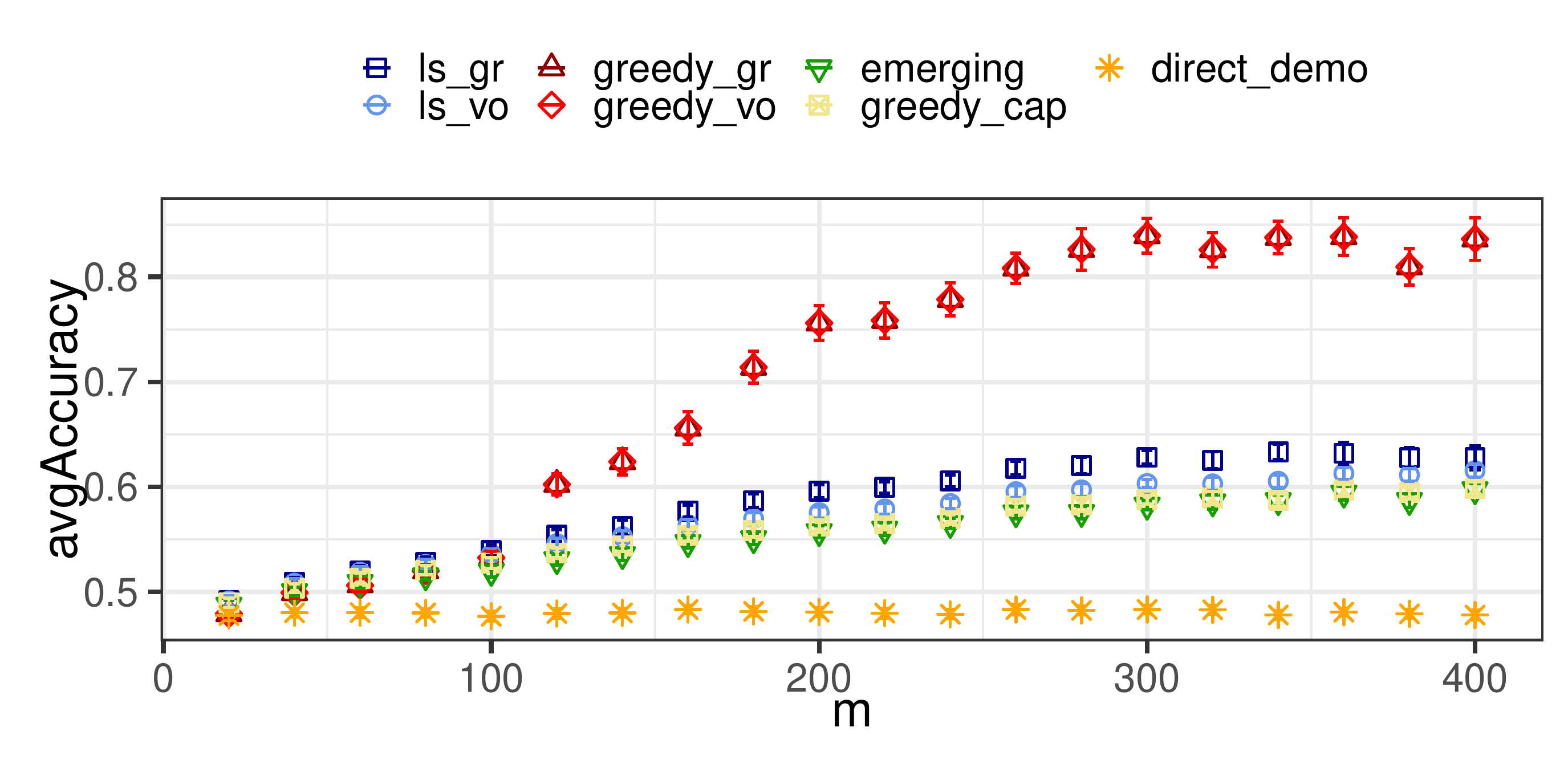}
	\includegraphics[trim=7mm 15mm 4mm 35mm, clip=true, width=.6\linewidth]{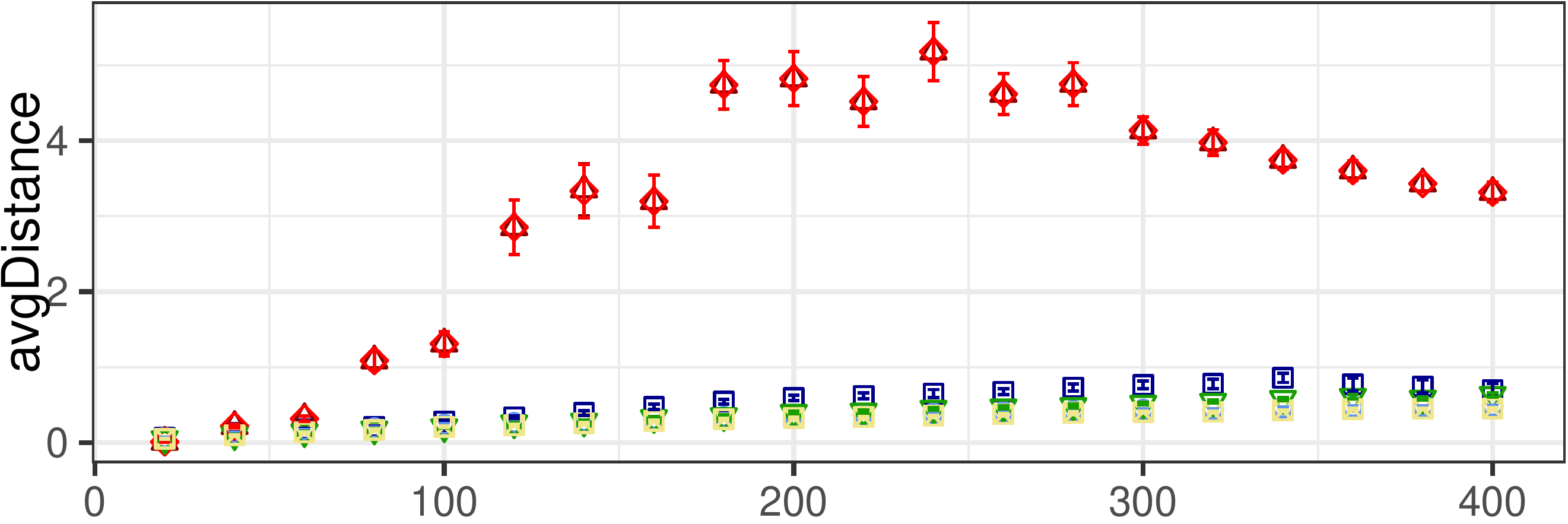}
	\includegraphics[trim=7mm 15mm 4mm 35mm, clip=true, width=.6\linewidth]{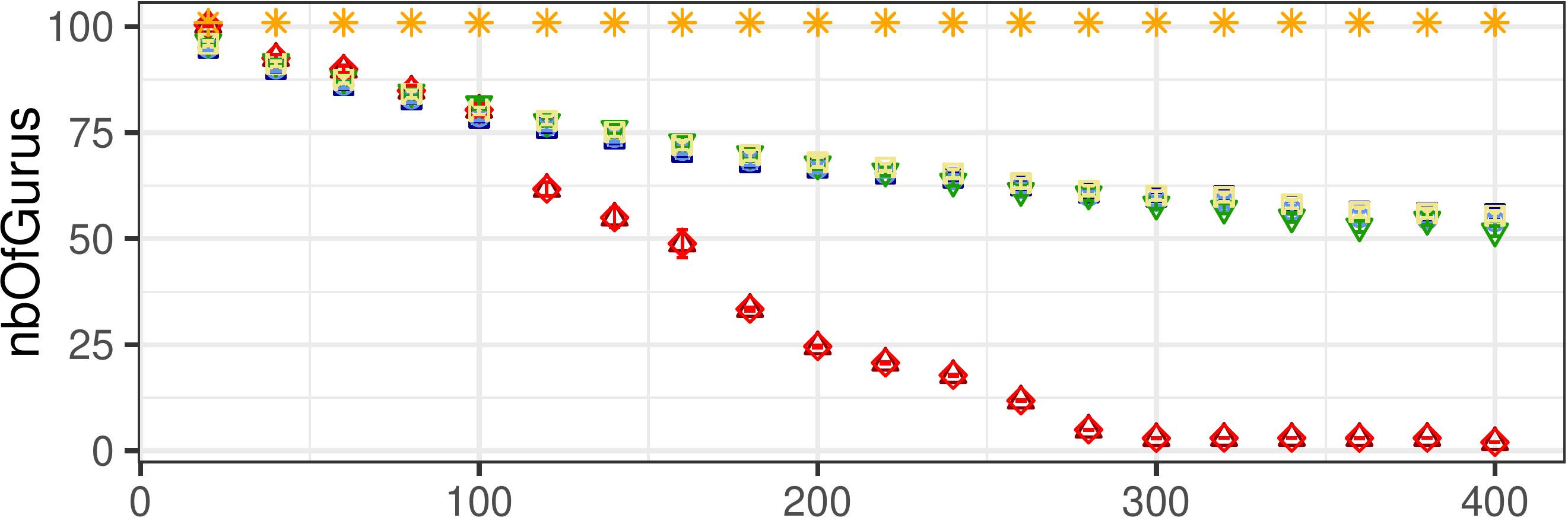}
	\includegraphics[trim=7mm  6mm 4mm 35mm, clip=true, width=.6\linewidth]{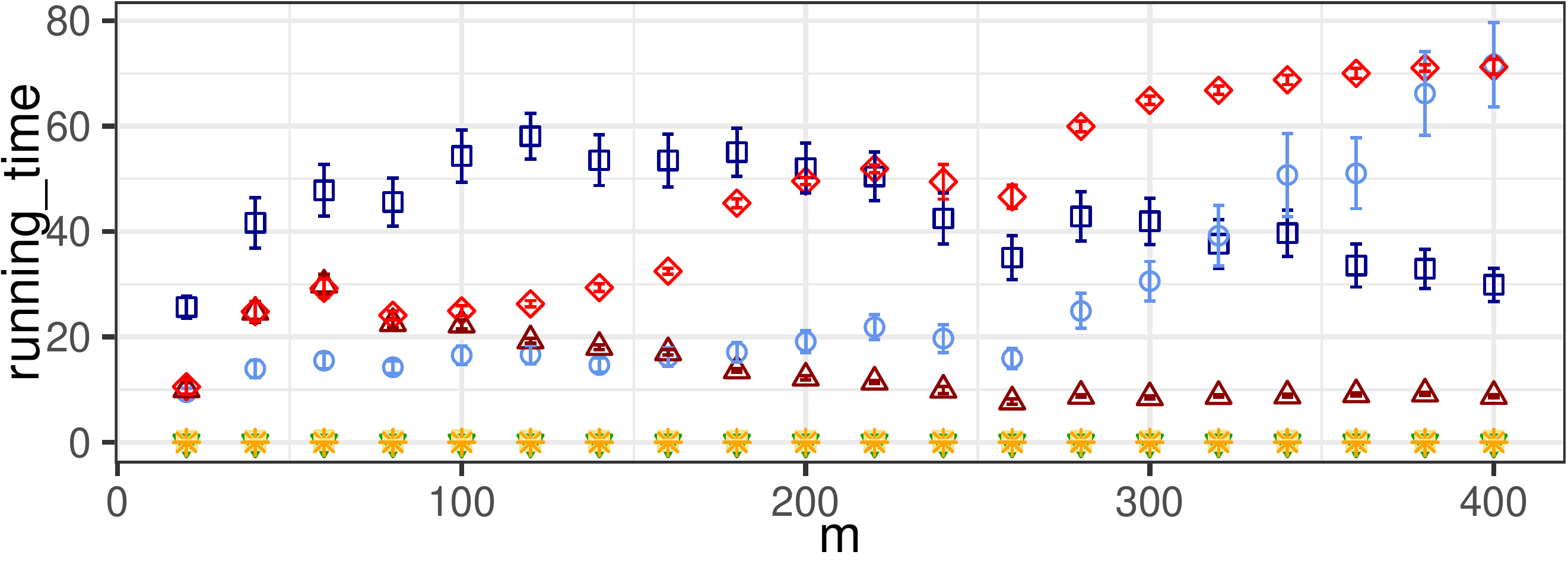}
	\caption{From top to bottom: average accuracies of gurus, average distance from voters to their guru, number of gurus, running time of the different methods using random $G_{n,m}$ graphs, $n = 101$ and m increasing from 20 to 400 in steps of 20 in all four plots. 
	}\label{fig : m}
\end{figure}

Figure~\ref{fig : m} shows the impact of increasing the number of edges $m = |E|$, i.e., the connectivity of the graph. We observe that this results in decreasing the number of gurus for all heuristics, while increasing the values of {\tt avgAccuracy} and {\tt avgDistance}. Note however, that at some point increasing $m$ would decrease the {\tt avgDistance} values again, once the graph is enough connected.

\begin{figure}[H]
	\centering
	\includegraphics[trim=7mm 15mm 4mm 10mm, clip=true, width=.6\linewidth]{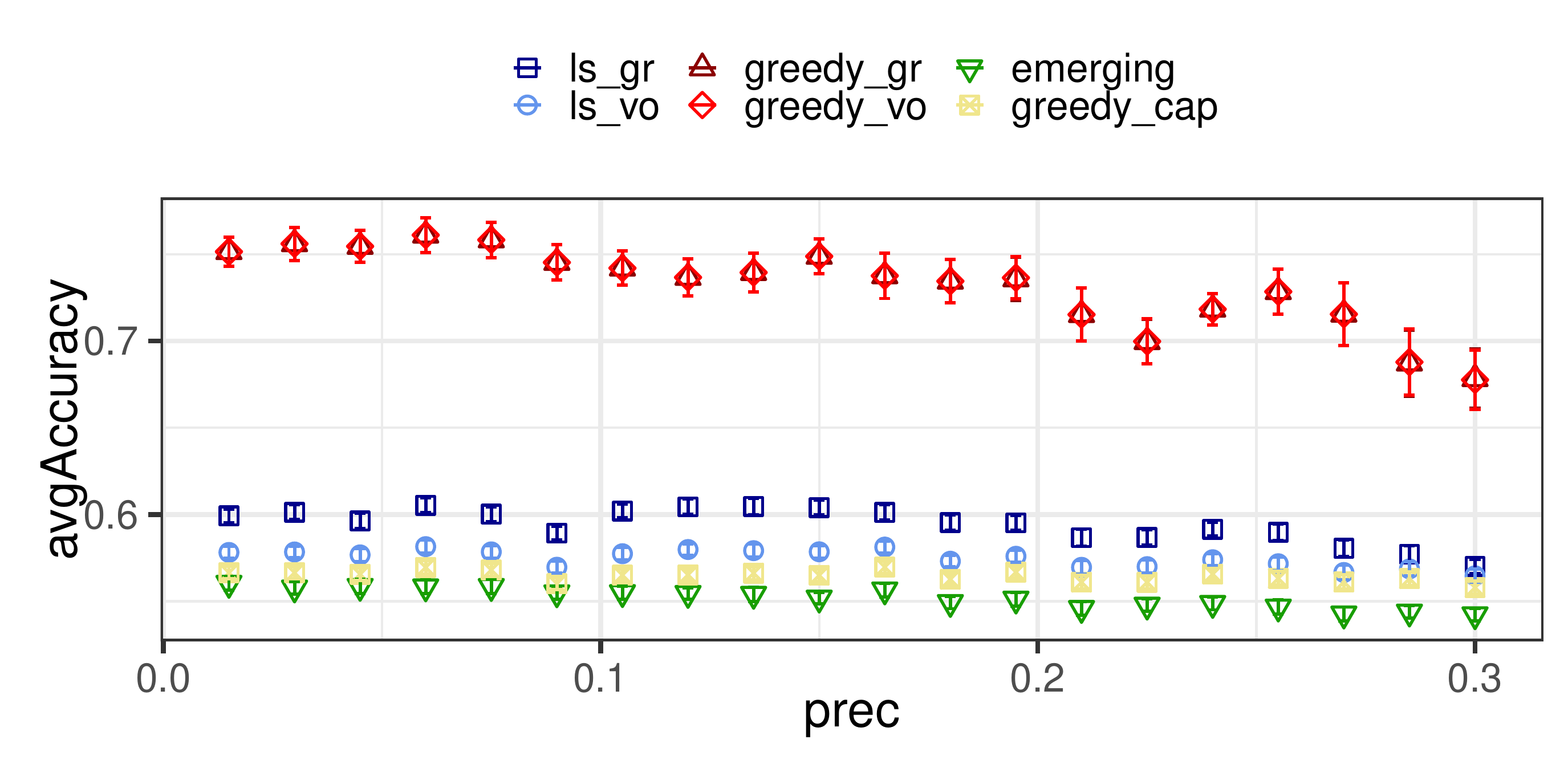}
	\includegraphics[trim=7mm 15mm 4mm 35mm, clip=true, width=.6\linewidth]{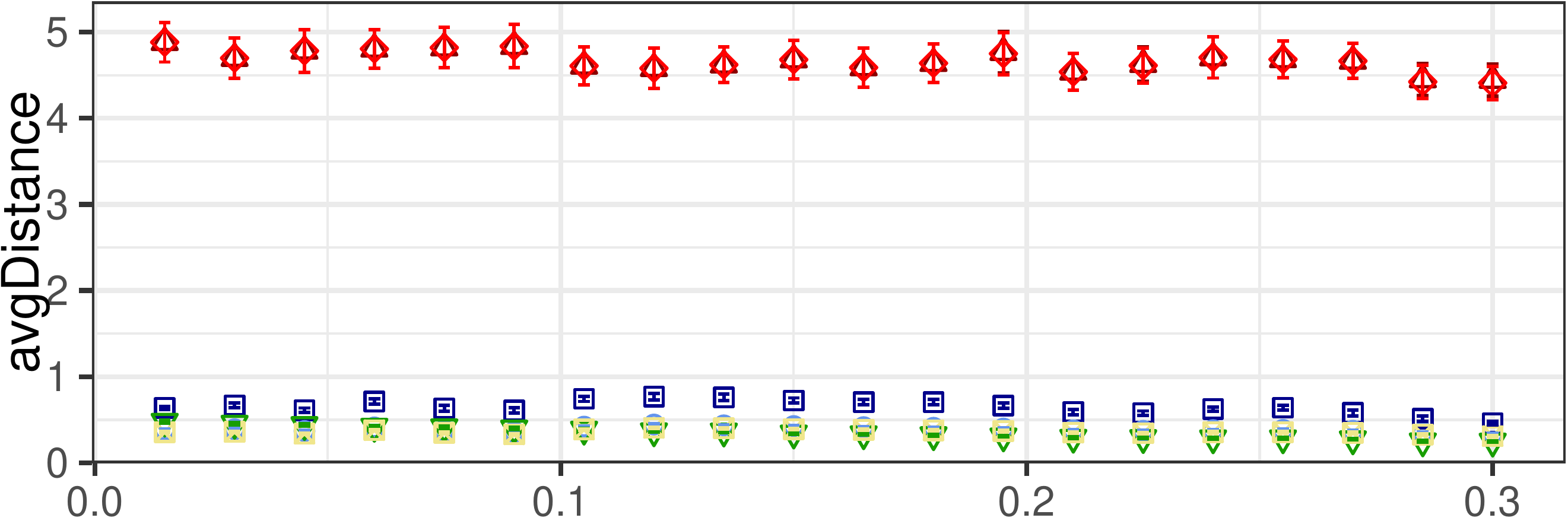}
	\includegraphics[trim=7mm 15mm 4mm 33mm, clip=true, width=.6\linewidth]{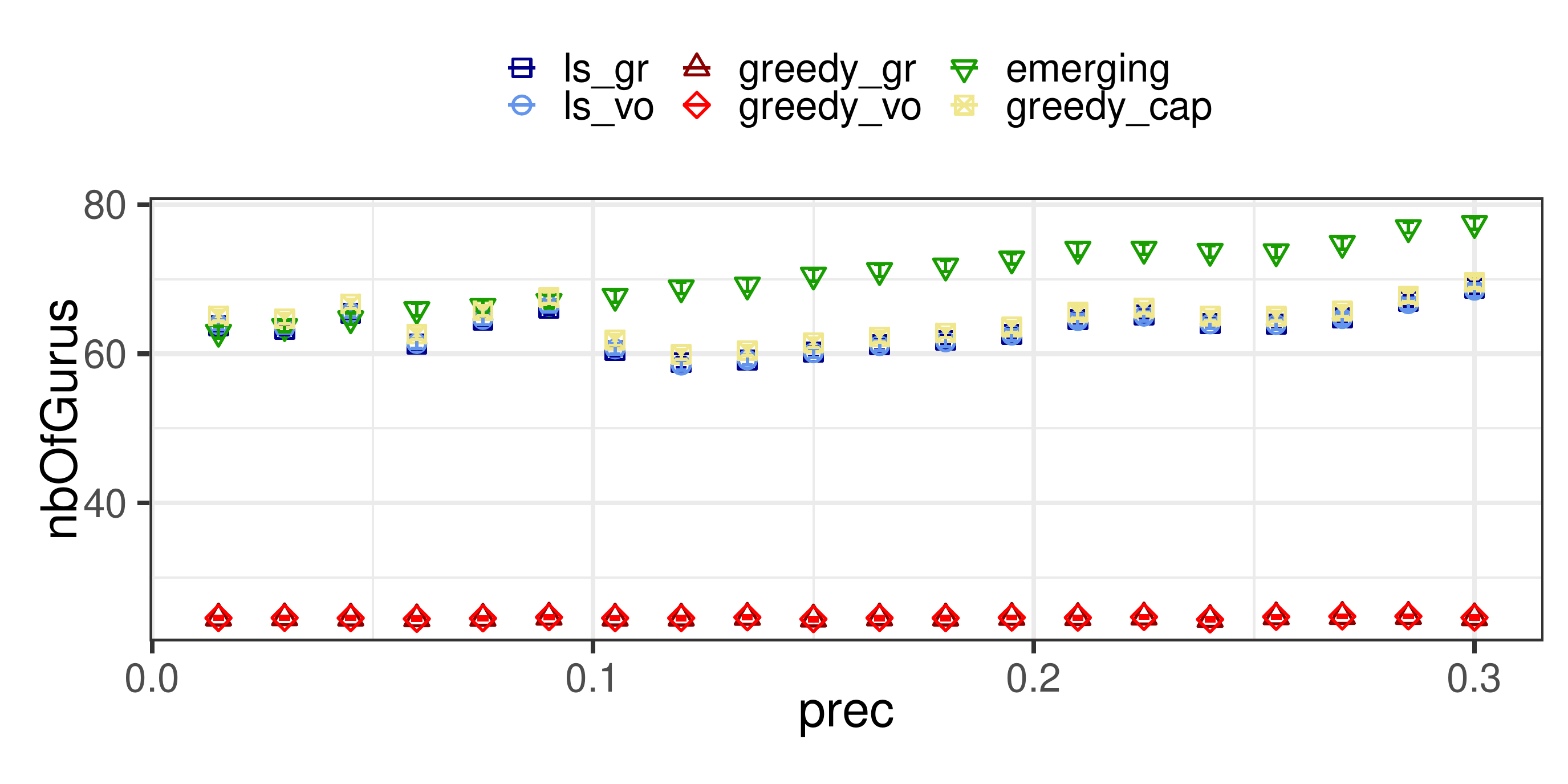}
	\includegraphics[trim=7mm  6mm 4mm 33mm, clip=true, width=.6\linewidth]{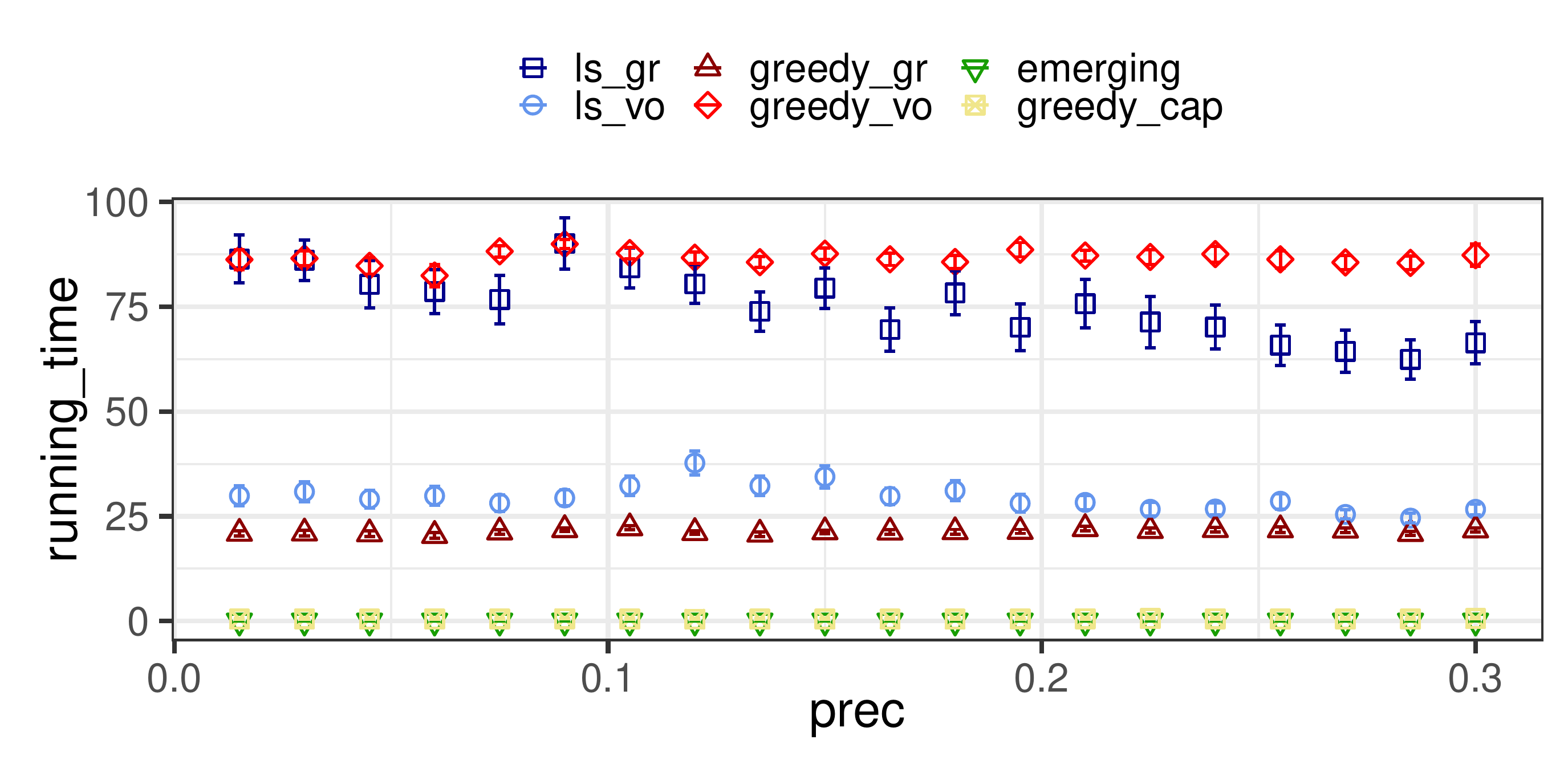}
	\caption{From top to bottom : average accuracies of gurus, average distance from voters to their guru, number of gurus, running time of the different methods using random $G_{n,m}$ graphs, $n = 101$ and $m = 2n$ and ${\tt prec}$ increases from $0.015$ to $0.3$ in steps of $0.015$ in all four plots. 
	} \label{fig : prec}
\end{figure}

Figure~\ref{fig : prec} shows the impact of increasing  parameter ${\tt prec}$. We observe that this results in increasing the number of gurus for all heuristics, while it decreases the value of {\tt avgAccuracy}.

\begin{figure}[H]
	\centering
	\includegraphics[trim=7mm 15mm 4mm 10mm, clip=true, width=.6\linewidth]{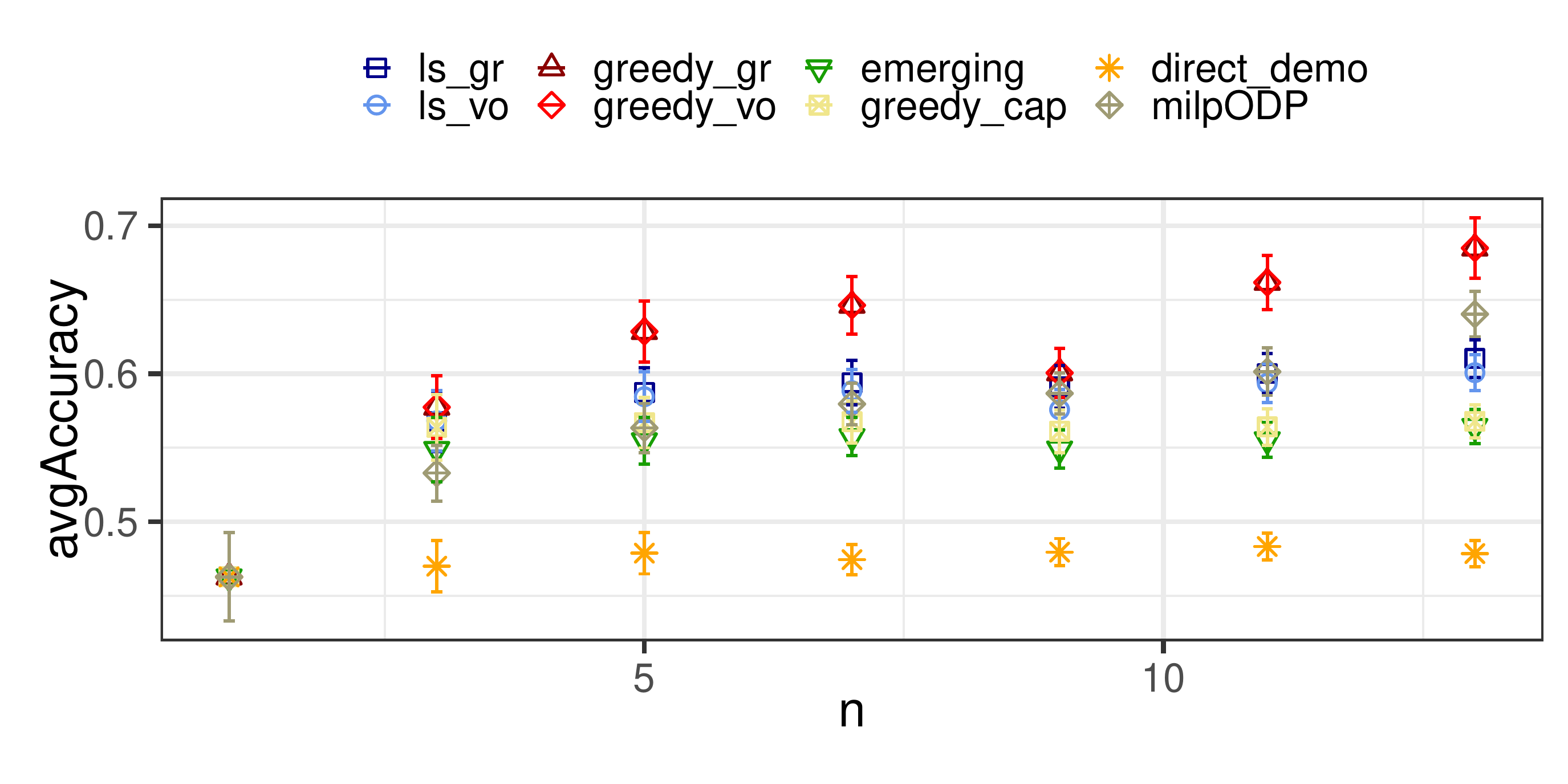}
	\includegraphics[trim=7mm 15mm 4mm 35mm, clip=true, width=.6\linewidth]{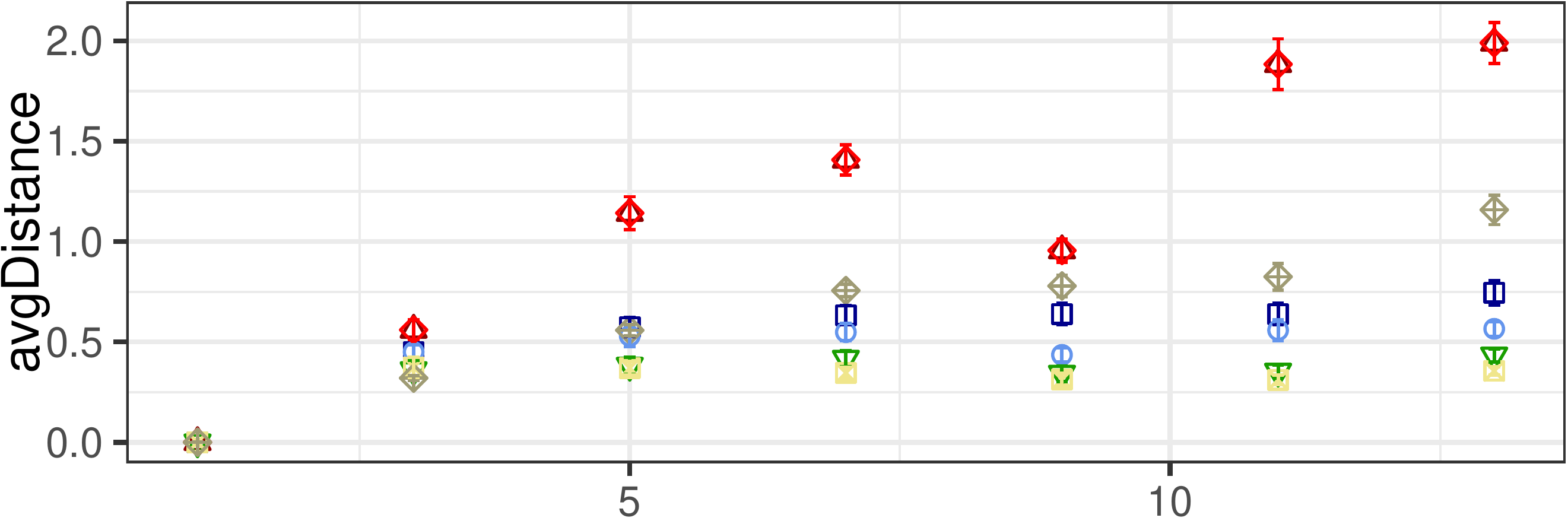}
	\includegraphics[trim=7mm 15mm 4mm 35mm, clip=true, width=.6\linewidth]{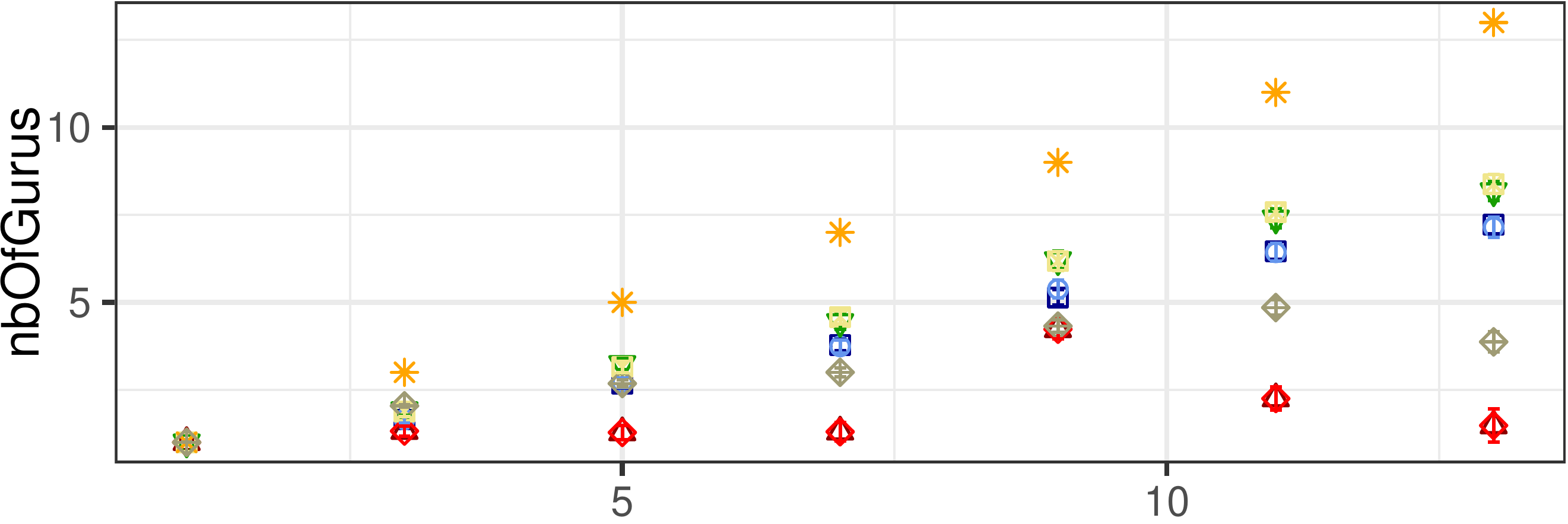}
	\includegraphics[trim=7mm  6mm 4mm 35mm, clip=true, width=.6\linewidth]{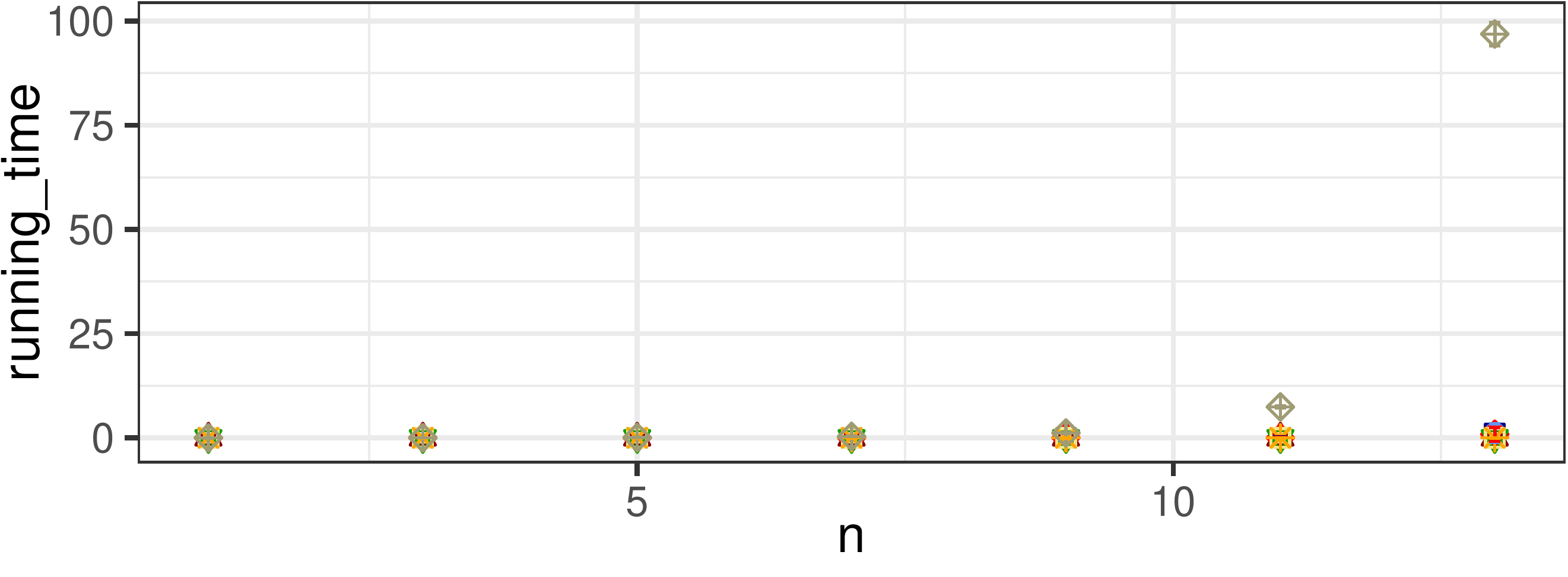}
	\caption{From top to bottom: average accuracies of gurus, average distance from voters to their guru, number of gurus, running time of the different methods using random $G_{n,m}$ graphs, $m=2n$ edges, $n$ increasing from 1 to 13 in steps of 2 in all four plots. 
	} \label{fig : exact}
\end{figure}

Figures~\ref{fig : exact} show that the features of an optimal solution returned by the MILP are often between the ones of the solutions returned by our greedy and local search strategies.

\clearpage

\end{document}